\documentclass[review,number,sort&compress]{elsarticle}
\usepackage{amssymb}
\usepackage{float}
\usepackage{siunitx}
\usepackage{subfigure}
\usepackage[mathlines]{lineno}
\usepackage[T1]{fontenc}
\DeclareSIUnit\curie{Ci}
\newcommand{\eps}{\epsilon}
\newcommand{\pce}{\eta^\varphi_\mathrm{NR}}
\newcommand{\geant}{G\textsc{eant}4~}
\newcommand{\oldgeant}{G\textsc{eant}3~}

\journal{Nuclear Instruments and Methods A}
\begin{document}
\begin{frontmatter}
\title{Nuclear-recoil energy scale in CDMS~II silicon dark-matter detectors}

\author[UFG]{R.~Agnese}\address[UFG]{Department of Physics, University of Florida, Gainesville, FL 32611, USA}
\author[mit]{A.J.~Anderson}\address[mit]{Department of Physics, Massachusetts Institute of Technology, Cambridge, MA 02139, USA}
\author[SLAC]{T.~Aramaki}\address[SLAC]{SLAC National Accelerator Laboratory/Kavli Institute for Particle Astrophysics and Cosmology, Menlo Park, CA 94025, USA}
\author[TAM]{W.~Baker}\address[TAM]{Department of Physics and Astronomy, and the Mitchell Institute for Fundamental Physics and Astronomy, Texas A\&M University, College Station, TX 77843, USA}
\author[SMU]{D.~Balakishiyeva}\address[SMU]{Department of Physics, Southern Methodist University, Dallas, TX 75275, USA}
\author[HBNI]{S.~Banik}\address[HBNI]{School of Physical Sciences, National Institute of Science Education and Research, HBNI, Jatni - 752050, India}
\author[UMN]{D.~Barker}\address[UMN]{School of Physics \& Astronomy, University of Minnesota, Minneapolis, MN 55455, USA}
\author[fermi,uiuc]{R.~Basu~Thakur}\address[fermi]{Fermi National Accelerator Laboratory, Batavia, IL 60510, USA}\address[uiuc]{Department of Physics, University of Illinois at Urbana-Champaign, Urbana, IL 61801, USA}
\author[fermi]{D.A.~Bauer}
\author[USD]{T.~Binder}\address[USD]{Department of Physics, University of South Dakota, Vermillion, SD 57069, USA}
\author[SLAC]{A.~Borgland}
\author[Sdsmt]{M.A.~Bowles\corref{cor1}}\address[Sdsmt]{Department of Physics, South Dakota School of Mines \& Technology, Rapid City, SD 57701, USA}
\ead{Michael.Bowles@Mines.sdsmt.edu, (315)5595864}
\cortext[cor1]{Corresponding author}
\author[SLAC]{P.L.~Brink} 
\author[PNNL]{R.~Bunker}\address[PNNL]{Pacific Northwest National Laboratory, Richland, WA 99352, USA}
\author[Stanford]{B.~Cabrera}\address[Stanford]{Department of Physics, Stanford University, Stanford, CA 94305, USA}
\author[UCSB]{D.O.~Caldwell$^{**}$\footnote{$^{**}$ Deceased}}\address[UCSB]{Department of Physics, University of California, Santa Barbara, CA 93106, USA}
\author[SMU]{R.~Calkins} 
\author[SLAC]{C.~Cartaro}
\author[UMadrid,Durham]{D.G.~Cerde\~no}\address[UMadrid]{Instituto de F\'{\i}sica Te\'orica UAM/CSIC, Universidad Aut\'onoma de Madrid, 28049 Madrid, Spain}\address[Durham]{Department of Physics, Durham University, Durham DH1 3LE, UK}
\author[Caltech]{Y.-Y.~Chang}\address[Caltech]{Division of Physics, Mathematics, \& Astronomy, California Institute of Technology, Pasadena, CA 91125, USA}
\author[UMN]{H.~Chagani}
\author[Syracuse,ualberta]{Y.~Chen} \address[Syracuse]{Department of Physics, Syracuse University, Syracuse, NY 13244, USA}\address[ualberta]{Department of Physics, University of Alberta, Edmonton, T6G 2E1, Canada}
\author[SMU]{J.~Cooley}
\author[Caltech]{B.~Cornell}
\author[UMN]{P.~Cushman}
\author[Berkeley]{M.~Daal}\address[Berkeley]{Department of Physics, University of California, Berkeley, CA 94720, USA}
\author[Berkeley]{T.~Doughty}
\author[CasePhy,CaseCS]{E.M.~Dragowsky}\address[CasePhy]{Department of Physics, Case Western Reserve University, Cleveland, OH 44106, USA}\address[CaseCS]{Research Computing, University Technologies, Case Western Reserve University, Cleveland, OH 44106, USA}
\author[UMadrid]{L.~Esteban}
\author[UMN,ualberta]{S.~Fallows}
\author[Queens]{E.~Fascione}\address[Queens]{Department of Physics, Queen's University, Kingston, ON K7L 3N6, Canada}
\author[NWestern]{E.~Figueroa-Feliciano}\address[NWestern]{Department of Physics \& Astronomy, Northwestern University, Evanston, IL 60208-3112, USA}
\author[UMN]{M.~Fritts} 
\author[Queens]{G.~Gerbier}
\author[Queens]{R.~Germond}
\author[Queens]{M.~Ghaith} 
\author[SLAC]{G.L.~Godfrey}
\author[Caltech]{S.R.~Golwala}
\author[SNOLAB]{J.~Hall}\address[SNOLAB]{SNOLAB, Creighton Mine \#9, 1039 Regional Road 24, Sudbury, ON P3Y 1N2, Canada}
\author[TAM]{H.R.~Harris} 
\author[fermi]{D.~Holmgren}
\author[NWestern]{Z.~Hong}
\author[fermi]{L.~Hsu}
\author[Denver,DenverEEP]{M.E.~Huber}\address[Denver]{Department of Physics, University of Colorado Denver, Denver, CO 80217, USA}\address[DenverEEP]{Departments of Physics and Electrical Engineering, University of Colorado Denver, Denver, CO 80217, USA}
\author[HBNI]{V.~Iyer}
\author[SMU]{D.~Jardin}
\author[TAM]{A.~Jastram}
\author[HBNI]{C.~Jena} 
\author[SLAC]{M.H.~Kelsey} 
\author[UMN]{A.~Kennedy} 
\author[TAM]{A.~Kubik}
\author[SLAC]{N.A.~Kurinsky}
\author[mit]{A.~Leder}
\author[Durham]{E.~Lopez~Asamar}
\author[fermi]{P.~Lukens}
\author[UBC,TRIUMF]{D.~MacDonell} \address[UBC]{Department of Physics \& Astronomy, University of British Columbia, Vancouver, BC V6T 1Z1, Canada}\address[TRIUMF]{TRIUMF, Vancouver, BC V6T 2A3, Canada}
\author[TAM]{R.~Mahapatra}
\author[UMN]{V.~Mandic}
\author[UMN]{N.~Mast}
\author[Diseased]{K.A.~McCarthy}\address[Diseased]{Institute for Disease Modeling, 3150 139th Ave SE, Bellevue, WA 98005, USA}
\author[Sdsmt]{E.H.~Miller}
\author[TAM]{N.~Mirabolfathi}
\author[Stanford]{R.A.~Moffatt}
\author[HBNI]{B.~Mohanty}
\author[Caltech,Yale]{D.~Moore}\address[Yale]{Department of Physics, Yale University, New Haven, CT 06520, USA}
\author[TAM]{J.D.~Morales~Mendoza}
\author[UMN]{J.~Nelson}
\author[UBC,TRIUMF]{S.M.~Oser}
\author[Queens]{K.~Page}
\author[UBC,TRIUMF]{W.A.~Page}
\author[SLAC]{R.~Partridge}
\author[Durham]{M.~Penalver~Martinez}
\author[UMN]{M.~Pepin}
\author[Berkeley]{A.~Phipps}
\author[USD]{S.~Poudel}
\author[Berkeley]{M.~Pyle}
\author[SMU]{H.~Qiu}
\author[Queens]{W.~Rau}
\author[Stanford]{P.~Redl}
\author[Evansville]{A.~Reisetter}\address[Evansville]{Department of Physics, University of Evansville, Evansville, IN 47722, USA}
\author[Denver]{A.~Roberts}
\author[UMN]{H.E.~Rogers}
\author[UMontreal]{A.E.~Robinson}\address[UMontreal]{Département de Physique, Université de Montréal, Montréal, Québec H3T 1J4, Canada} 
\author[UFG]{T.~Saab}
\author[Berkeley,LBNL]{B.~Sadoulet}\address[LBNL]{Lawrence Berkeley National Laboratory, Berkeley, CA 94720, USA}
\author[USD]{J.~Sander}
\author[SLAC]{K.~Schneck}
\author[Sdsmt]{R.W.~Schnee}
\author[SNOLAB]{S.~Scorza}
\author[HBNI]{K.~Senapati}
\author[Berkeley]{B.~Serfass}
\author[Berkeley]{D.~Speller}
\author[Queens]{P.C.F.~Di~Stefano} 
\author[SMU]{M.~Stein}
\author[Sdsmt]{J.~Street}
\author[UToronto]{H.A.~Tanaka}\address[UToronto]{Department of Physics, University of Toronto, Toronto, ON M5S 1A7, Canada}
\author[TAM]{D.~Toback}
\author[Queens]{R.~Underwood}
\author[UMN,Denver]{A.N.~Villano}
\author[UBC,TRIUMF]{B.~von~Krosigk}
\author[UFG]{B.~Welliver}
\author[TAM]{J.S.~Wilson}
\author[UToronto]{M.J.~Wilson}
\author[SLAC]{D.H.~Wright}
\author[Stanford]{S.~Yellin}
\author[Stanford]{J.J.~Yen}
\author[SantaClara]{B.A.~Young}\address[SantaClara]{Department of Physics, Santa Clara University, Santa Clara, CA 95053, USA}
\author[Queens]{X.~Zhang}
\author[TAM]{X.~Zhao}

\newpage
\begin{abstract}
{\small
The Cryogenic Dark Matter Search (CDMS~II) experiment aims to detect 
dark matter particles that elastically scatter from 
nuclei in semiconductor detectors. 
The resulting nuclear-recoil energy depositions 
are detected by ionization and phonon sensors. 
Neutrons produce a similar spectrum of low-energy nuclear recoils 
in such detectors, while most other backgrounds produce electron recoils. 
The absolute energy scale for nuclear recoils 
is necessary to interpret results correctly.
The energy scale can be determined in CDMS~II silicon detectors  
using neutrons incident 
from a broad-spectrum $^{252}$Cf source, taking advantage 
of a prominent resonance in the neutron elastic scattering cross section 
of silicon at a recoil (neutron) energy near 20 (182)\,keV.
Results indicate that the phonon collection efficiency for nuclear recoils is 
$4.8^{+0.7}_{-0.9}$\% lower than for electron recoils of the same energy. 
Comparisons of the ionization signals for nuclear recoils to those 
measured previously by other groups at higher electric fields  
indicate that the ionization collection efficiency for CDMS~II silicon detectors 
operated at $\sim$4\,V/cm is 
consistent with 100\% for nuclear recoils below 20\,keV 
and gradually decreases for larger energies to $\sim$75\% at 100\,keV. 
The impact of these measurements on previously published CDMS~II silicon results is small.
}
\end{abstract}
\newpage

\end{frontmatter}


\section{Introduction}

Strong evidence indicates that $\gtrsim$80\% 
of the matter in the Universe is non-luminous and non-baryonic~\cite{FengReview}.  
Weakly Interacting Massive Particles (WIMPs) are a leading class 
of theoretically motivated candidates for this dark matter~\cite{steigmanturner}. 
These particles are expected to interact with normal matter 
through the weak nuclear force and to cluster gravitationally. 
If WIMPs do constitute our galaxy's dark matter, 
they may be detectable through their elastic scattering off 
atomic nuclei in terrestrial detectors~\cite{goodmanwitten}. 
Under standard galactic halo assumptions~\cite{lewinsmith} 
for a WIMP mass of $\sim$\SI{100}{\GeV}/$c^2$, 
the recoiling nuclei have energies of tens of \si{\keV} and 
ranges of 10\textendash100 nm in solid matter.

The Cryogenic Dark Matter Search (CDMS~II) experiment 
measured nuclear recoils using a target mass 
composed of high-purity silicon and germanium 
semiconductor crystals operated at $\sim$\SI{50}{\milli\K}.  
Each crystal was instrumented to simultaneously measure 
the electron-hole pairs (ionization) 
and athermal phonons created by particle interactions within the crystal~\cite{r21lowmass}.

A WIMP, or a neutron, may scatter off a nucleus  
producing a nuclear recoil (NR), 
while most other interactions produce an electron recoil (ER). 
Accurate determination of an event's energy 
requires a systematic calibration of the recoil energy scale.
This energy calibration is generally straightforward 
for electron recoils due to the 
availability of a variety of spectral lines from 
radioactive sources over a wide range of energies. 

The calibration for nuclear recoils is more difficult. 
CDMS~II used a  $^{252}$Cf neutron source to perform nuclear-recoil calibrations, 
and the spectrum of recoil energies in CDMS~II detectors 
resulting from exposure to this source 
decreases quasi-exponentially with increasing energy and is nearly featureless.
For CDMS~II detectors,
knowledge of the nuclear-recoil energy scale to within $\sim$10\% 
is sufficient to accurately interpret WIMP-search results 
for WIMP masses greater than a few tens of GeV/$c^2$. 
For lower masses, however, a more accurate determination of
the energy scale is important for a robust comparison 
of results from different experiments, particularly
in light of interpretations of 
data from several experiments as possible evidence 
for a low-mass (<\SI{10}{\GeV}/$c^2$) 
WIMP~\cite{dama2010,cogent_prl_2011, cresstii_2012_EPJC, si_CDMSII_PRL}. 

This paper describes the procedure used to calibrate 
the nuclear recoil response of CDMS~II silicon detectors. 
Experimental data for this study are drawn from the 
final runs of these detectors at the Soudan Underground Laboratory, 
from July~2007 to September~2008,  
as described in Ref.~\cite{si_CDMSII_PRL}. 
{\it In situ} measurements of elastic neutron scatters in these detectors from a $^{252}$Cf source are compared to Monte Carlo simulations of  recoiling nuclei in the detectors.
A re-calibrated energy scale is derived, 
optimizing agreement between measured and simulated recoil spectra. 
This is used to adjust the published upper limits on the WIMP-nucleon spin-independent cross section, as well as the 90\%~C.L. acceptance region from the analysis of the final exposure of the silicon detectors~\cite{si_CDMSII_PRL}.

\section{CDMS~II Detectors}
The final configuration of CDMS~II contained 
11~silicon and 19~germanium Z-sensitive Ionization- and Phonon-mediated (ZIP) detectors. 
These were arrayed 
into five ``towers,'' 
each containing six detectors
following the designation T$x$Z$y$  where $x$ (1\textendash5) 
is the tower number and $y$ (1\textendash6) indicates the 
position within the stack (from top to bottom).
We focus here on the silicon detectors used in Ref.~\cite{si_CDMSII_PRL}, 
which were $\sim$\SI{10}{\mm} thick, \SI{76}{\mm} in diameter, 
with a mass of $\sim$\SI{106}{\g} each. 
Of the eleven silicon detectors, 
two were excluded due to wiring failures leading to incomplete ionization collection, 
and a third was excluded due to unstable phonon channel response. 

Each detector was photolithographically patterned with sensors on both flat faces: 
two concentric ionization electrodes on one face and four independent phonon sensors on the opposite face. 
The ionization electrodes were biased to \SI{4}{\V} with respect to the phonon electrodes, 
creating an electric field of 4 V/cm in the bulk of the detector along its $z$ axis~\cite{cdms_gePosCorr_PRD}. 
The electrons and holes generated by a particle interaction were separated and drifted across the crystal by the electric field, 
generating image currents in the electrodes 
detected by a JFET-based charge amplifier~\cite{cdmsreadout}. 
By careful neutralization of ionized trapping sites within the crystal with regular exposure to infrared LEDs (``flashing''), 
the detectors were operated in a metastable state in which trapping of charge carriers in the crystal bulk was low. 
The ionization collection efficiency for electron recoils was therefore high, despite the relatively modest applied electric field.

In semiconductor devices such as the ZIPs, phonon ($\varphi$) energy is generated by three interactions: 
the initial recoil generates primary phonons, 
the work done on the charge carriers by the electric field 
generates Neganov-Trofimov-Luke (or NTL) phonons~\cite{Neganov:1985khw, luke, neganov}, 
and charge carrier relaxation to the Fermi level at the electrodes 
generates recombination phonons.
When a particle interacts in a ZIP, 
it deposits a recoil energy $E_{\rm R}$ in the crystal and generates $n_Q$ electron-hole pairs. 
For electron recoils, this recoil energy $E_\textrm{R} = n_{Q}\epsilon$, 
where $\eps$ is the average energy required to create one electron-hole 
pair.\footnote{For silicon, $\eps=3.82$\,eV above \SI{77}{\K}~\cite{ogburnthesis} 
and is not expected to deviate significantly at lower temperatures.}
A portion of this energy is stored in the potential energy of the 
drifting charge carriers and is restored to the phonon system when 
they relax to the Fermi level at the electrodes, 
producing recombination phonons.

The work done by the electric field on the $n_Q$ drifting charge 
pairs results in the Cherenkov-like radiation of an additional population 
of phonons at near-ballistic energies. 
These are the so-called NTL phonons which add a contribution 
to the total phonon signal proportional to the bias voltage $V_{\mathrm b}$ across the detector:  
$E_\textrm{NTL} = n_Q e V_{\mathrm b}$. 
The total phonon energy 
is therefore $E_\varphi = E_{\rm R} + n_Q e V_{\mathrm b}$.  
It is convenient to express the ionization signal as an 
electron-equivalent energy $E_Q \equiv n_Q \eps $ 
and the total phonon energy as
\begin{equation}
E_\varphi = E_{\rm R} + E_Q \frac{e V_{\mathrm b}}{\epsilon} = E_{\rm R} \left( 1 + y \frac{e V_{\mathrm b}}{\epsilon}\right),
\label{eqn:PhononEnergy}
\end{equation}
where $y \equiv E_Q / E_{\rm R}$ is the ionization yield.
With these definitions, an ideal electron recoil has ionization yield $y=1$.
An event's recoil energy is determined by rearranging Eqn.~\ref{eqn:PhononEnergy}, 
$E_{\rm R } = E_\varphi - E_Q (e V_{\mathrm b}/\epsilon),$  where $E_\varphi$ is estimated from the phonon channels 
and $E_Q$ from the charge channels.

The remainder of the recoil energy is deposited directly into 
the phonon system as primary phonons. 
These high-frequency phonons undergo isotopic scattering and 
cannot travel far from their production sites before 
down-converting via anharmonic decay~\cite{ltd14AndersonTES} 
into lower-frequency phonons with larger mean free paths, 
comparable to the thickness of the detector~\cite{hertelthesis}.
The lower-frequency ballistic phonons then interact with either 
the phonon sensors or un-instrumented material at the detector surfaces. 

Details of the phonon collection mechanism in CDMS detectors are discussed in Ref.~\cite{lemanMCpaper}. 
Past analyses assumed that all 
three phonon contributions are detected with equal efficiency. 
This is a plausible assumption because all three mechanisms 
generally inject energy into the phonon system above the ballistic propagation threshold. 
All three types down-convert until they become just barely ballistic,  
so their frequency distributions at the sensors are nearly the same. 
However, the relative fraction of phonons absorbed by the 
sensors (compared to other materials) may depend on details of the 
primary interaction, and even on the relative fractions of primary and NTL phonons. 
Consequently, although the differences are expected to be small, 
the phonon collection efficiency 
in CDMS~II detectors 
for nuclear and electron recoils of a given energy need not be identical.
This paper describes measurements of the small difference between these two efficiencies.

\subsection{Electron-Recoil Calibration}\label{sec:electronRecoilCalib}
The response of the ZIP detectors to phonons and ionization 
from electron recoils is calibrated {\it in situ} using a gamma-ray source. 
Event-selection cuts are applied to electron-recoil 
calibration data to remove events with pathologies, 
including electronic glitch events, 
anomalously shaped charge pulses, 
and periods of high baseline noise.
From this sample, only those events  
occurring within a detector's fiducial volume (or ``bulk'') are selected, 
thereby avoiding surface events, which can suffer from incomplete ionization collection.
We reject events outside the detector's bulk 
by requiring the signal in the outer ionization electrode 
be consistent with noise, while the inner ionization signal 
must exceed a detector- and run-dependent threshold 4.5$\sigma$ 
above the noise mean (as described in Ref.~\cite{si_CDMSII_PRL}).

\subsubsection{Ionization calibration of electron recoils}\label{sec:ioniz_calib_e-}

A $^{133}$Ba gamma source with spectral lines at 275, 303, 356, and \SI{384}{\keV} 
was used to calibrate the ionization energy scale in the detectors. 
A significant number of these gamma rays are fully contained within a germanium detector,
producing clear peaks in histograms of the ionization pulse amplitude. 
The reconstructed ionization pulse amplitude 
from the germanium detectors is thus calibrated to an 
electron-equivalent recoil energy (keV$_{\textrm{ee}}$) 
by multiplying by a constant factor chosen 
such that the observed peaks lie at the appropriate Ba-line energies. 

\begin{figure}
\centering
\includegraphics[width=0.7\linewidth]{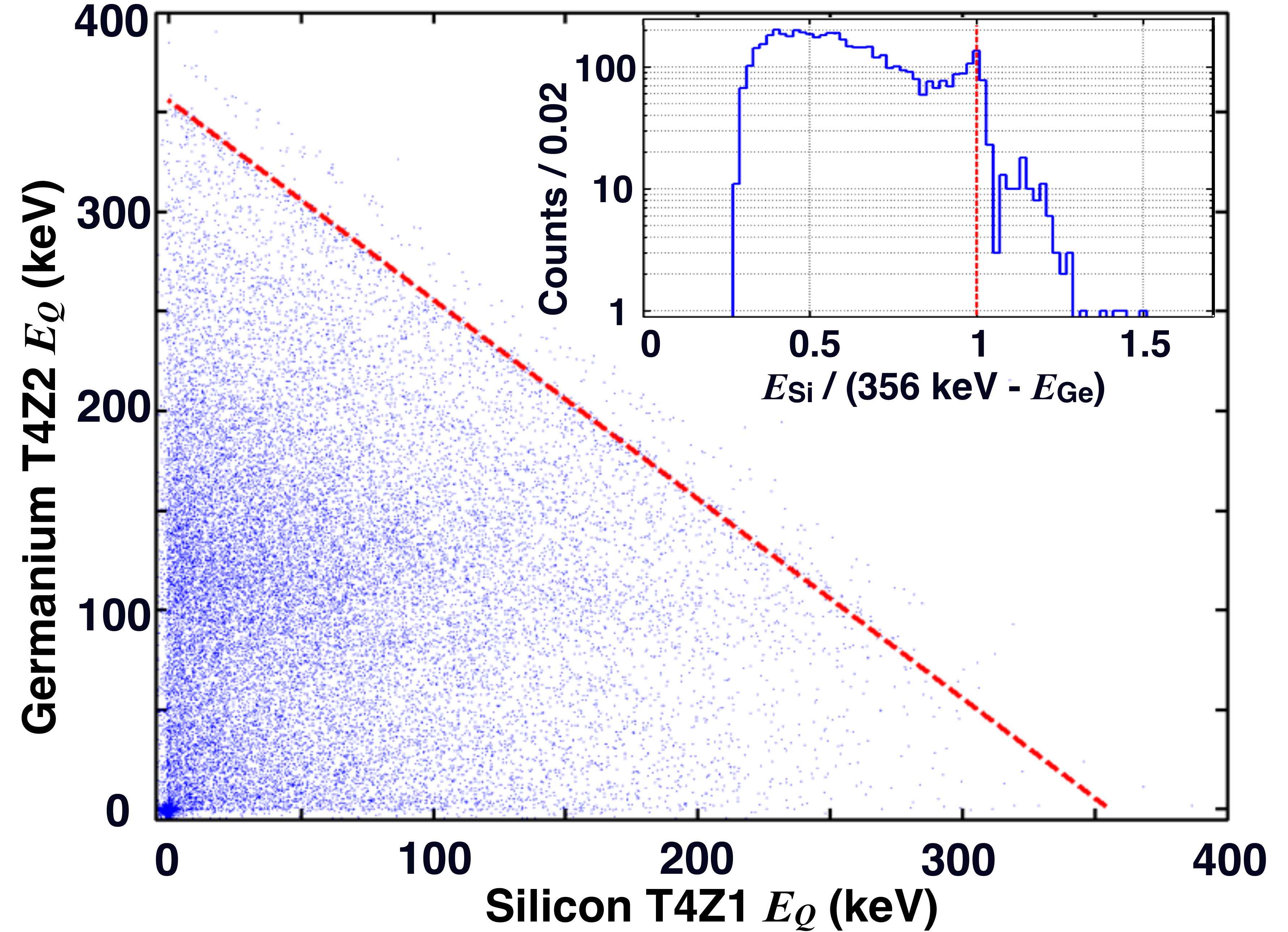}
\caption{Calibration of a silicon detector's ionization energy scale 
using the ionization collected from $^{133}$Ba gamma rays that deposited energy 
in both the silicon detector (T4Z1) and an adjacent germanium detector (T4Z2).  
The scatter plot shows the ionization energy $E_{Q}$ in the neighboring germanium detector as 
a function of the silicon-detector ionization energy. Events for which 
the full energy of $^{133}$Ba 356~keV gamma rays is deposited in 
the detector pair follow a diagonal feature (dashed line), 
enabling calibration of the silicon-detector energy scale and 
demonstrating linearity of the silicon-detector ionization response up to $>$\SI{350}{\keV}. 
{\bf Inset:} Same data histogrammed (with bin width~0.02) to show the ratio of the 
silicon-detector ionization energy to the expected 356\,keV gamma-ray 
energy less the germanium-detector ionization energy. 
A peak is clearly visible (dashed line) corresponding 
to 356~keV gamma rays that are fully contained by the detector pair. }
\label{fig:Si_Calibration}
\end{figure}

Because of their relatively low stopping power,
silicon detectors of this size rarely contain the full energy of 
the $^{133}$Ba gamma rays,  
so the peaks are not visible.
Silicon also has no intrinsic spectral lines at energies below \SI{100}{\keV}. 
The ionization energy scales in the silicon detectors 
are therefore calibrated using 
shared events---a \SI{356}{\keV} $^{133}$Ba gamma ray that 
deposits its energy within adjacent detectors. 
The $^{133}$Ba spectral lines are clearly visible in the 
sum of ionization energy $E_Q$ in a silicon detector 
and its germanium neighbor, as shown in Fig.~\ref{fig:Si_Calibration}.
Ionization energy scales are calibrated first for germanium detectors, 
and the calibration for silicon detectors is then set 
so that the shared event lines lie at the appropriate energies.

After confirming linearity in the germanium detectors across a 
wide range of spectral lines, 
linearity in the silicon detectors is checked
implicitly by tracking the total energy of 
shared \SI{356}{\keV} events as a function
of the reconstructed ionization energy in the silicon detector.
The position of this peak varies by less than 5\%, 
demonstrating linearity up to $>$\SI{350}{\keV}.

\subsubsection{Phonon Calibration}

\begin{figure}[th]
\centering
\includegraphics[width=0.65\linewidth]{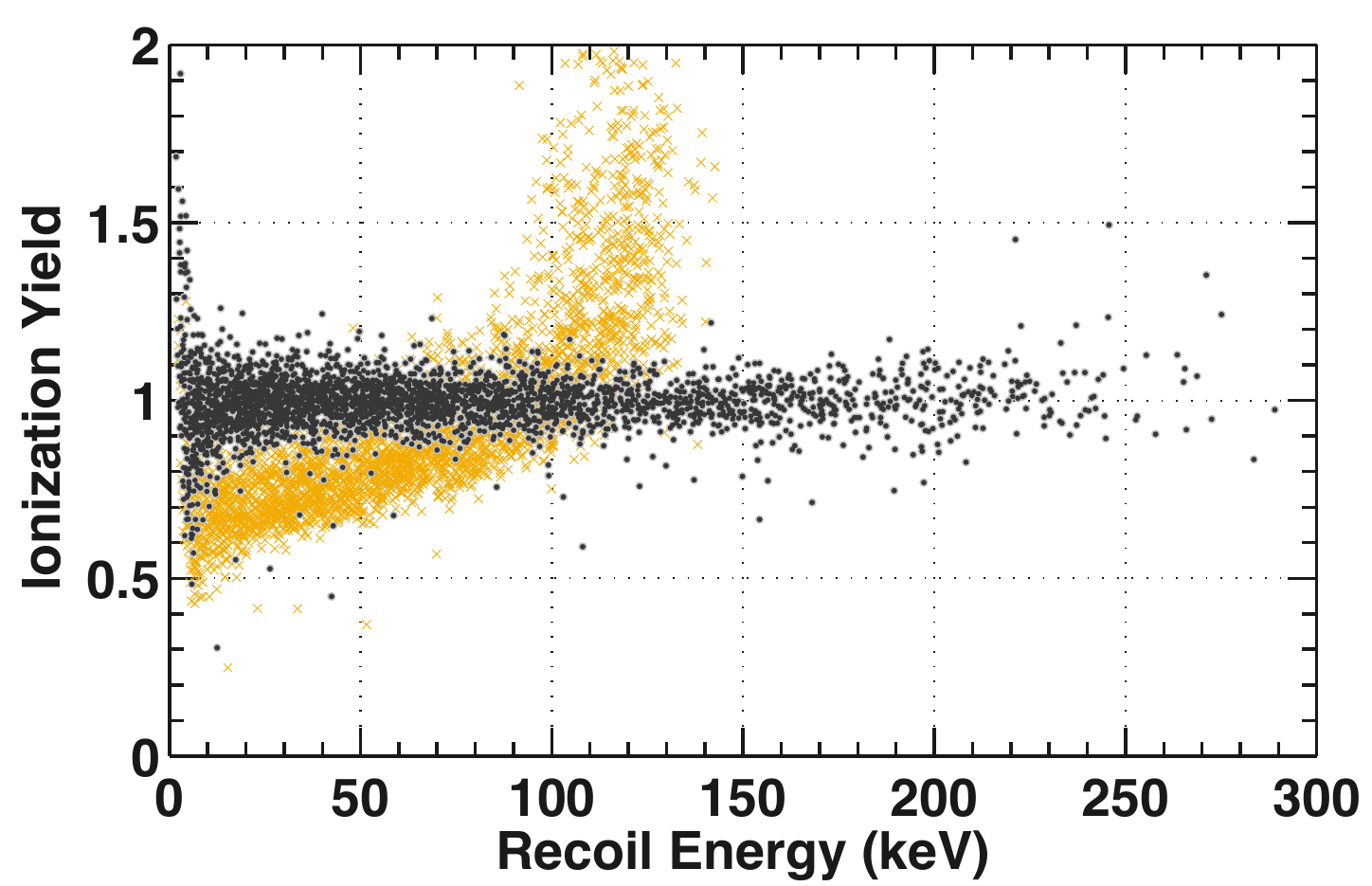}  
\caption{Ionization yield versus recoil energy for 
$^{133}$Ba calibration events in silicon detector T2Z2, 
both prior to applying the position-correction table 
to the phonon signal (light $\times$) and after (dark $\bullet$).\label{fig:2dposcorr}}
\end{figure} 
The energy scale for the phonon channels is calibrated 
using a sample of bulk electron recoils, 
which should have unity ionization yield; 
the reconstructed amplitude of the total phonon pulse is scaled 
so that the inferred recoil energy matches the ionization energy and thus gives $y = 1$ (on average).
The measured phonon signals have a significant 
position dependence that is removed in this process.
Based on position-reconstruction parameters  
derived from the relative amplitudes and timings of the
four phonon sensor signals, the broad continuum of
$^{133}$Ba electron recoils is used to develop an 
empirical correction table as a function of 
position, amplitude, and phonon energy
(as was done for the germanium detectors in Ref.~\cite{cdms_gePosCorr_PRD}). 
Figure~\ref{fig:2dposcorr} shows how application of this 
position-correction table removes the energy dependence (and improves the resolution) of the ionization yield for electron recoils.

\section{Constraining the Energy Scale for Nuclear Recoils}
\label{sec:NRcalibration}

\begin{figure}
\centering
\includegraphics[width=0.59\linewidth]{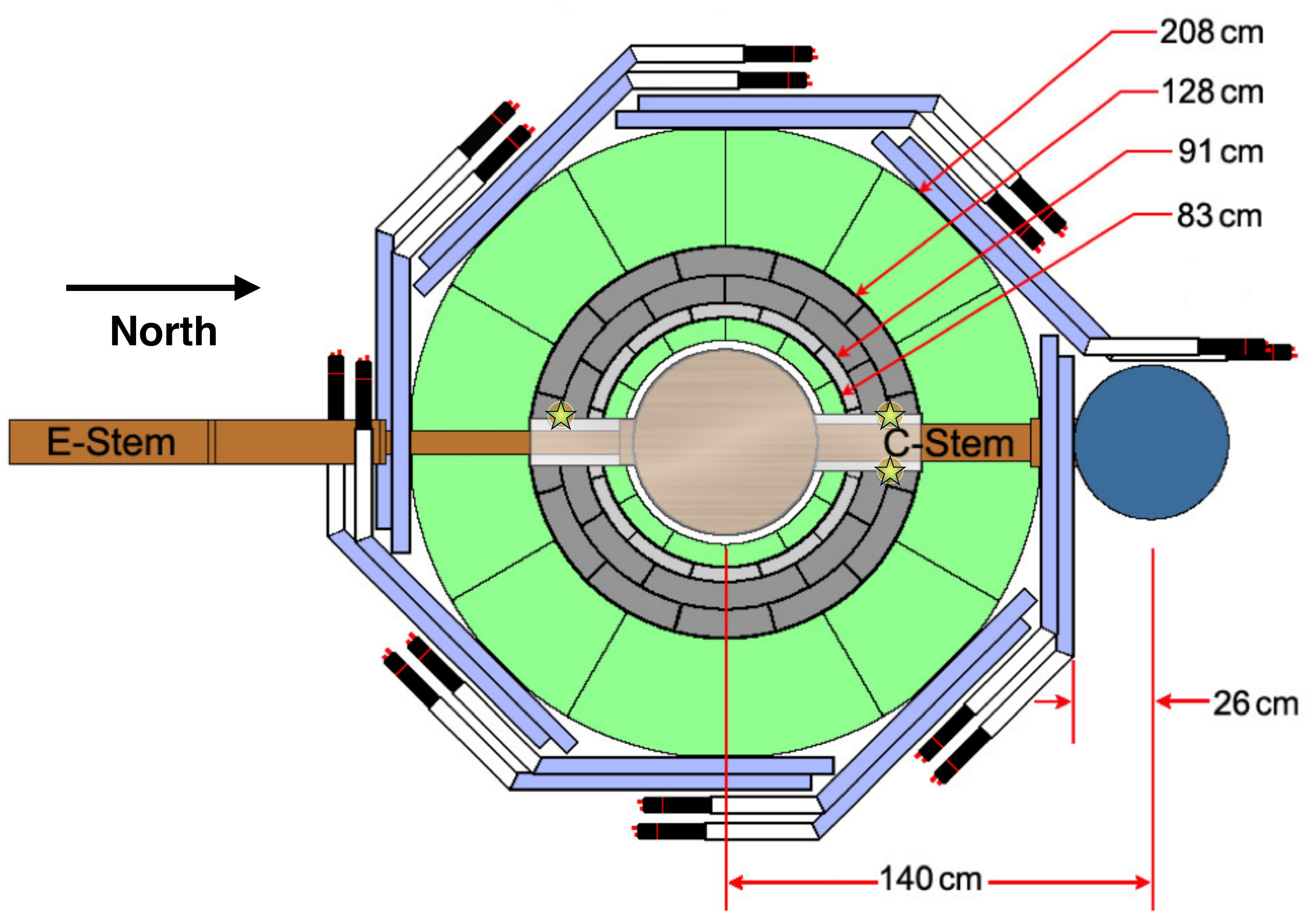}
\caption{(Color online) Top view of the CDMS~II apparatus 
with calibration source locations 
(northwest [NW], southwest [SW], and northeast [NE]) 
indicated by $\bigstar$. 
The muon veto panels are shown as the outermost, staggered layers (light blue), 
surrounding the outer annular layer of polyethylene (green), 
followed by a layer of low-radioactivity lead bricks (gray), 
a thin inner layer of ancient lead (light gray), 
an inner polyethylene shield (green), 
and finally a mu-metal shield (transparent gray) 
surrounding the copper cryostats cans (bronze). 
The mu-metal shield extends into two penetrations 
that pass through all layers of shielding to enable connections 
to the electronics readouts (`E-Stem') 
and the dilution refrigerator (`C-Stem'). 
\label{fig:calgeo}}
\end{figure}

Nuclear recoils were provided by a \SI{5}{\micro\curie} 
$^{252}$Cf neutron  source. 
Neutron capture causes temporary activation of the germanium detectors,   
so this calibration was performed less frequently than the $^{133}$Ba gamma-ray calibration. 
The period considered here contains six sequences of neutron calibration. 
During each of these sequences, 
several data sets were acquired 
with the source inserted into one of 
three plastic tubes running along straight paths 
through the polyethylene and lead shielding to within 10~cm 
of the  copper cryostat cans that housed the detectors, 
as shown in Fig.~\ref{fig:calgeo}. 
Each tube was labeled by its nearest inter-cardinal 
direction: southwest, northwest, or northeast.
Because each source position illuminated the detectors 
with a different relative neutron flux, 
calibration data were grouped by position 
and the resulting spectra were normalized separately.
Recoil energies for these events were calculated 
by subtracting the NTL phonon contribution, 
inferred from the ionization signal, 
from the total phonon energy.
However, unlike the $^{133}$Ba data, 
neutron calibration data have no clear spectral lines.
The resulting nuclear-recoil energy scale cannot 
be directly verified for correctness or linearity.

Instead, a \geant Monte Carlo simulation was performed 
with the goal of finding the linear energy scaling factor 
$\pce$---interpreted as the phonon collection efficiency 
of nuclear recoils relative to that of electron recoils---that 
minimizes a test statistic comparing 
the simulated spectra of nuclear recoil energies to the measured spectra.
The simulation geometry corresponded to the full experimental apparatus in the five-tower configuration used for CDMS~II, 
including the detectors, support structure, and all shielding. 
A $^{252}$Cf source was simulated separately at the three locations depicted in Fig.~\ref{fig:calgeo}.
Neutrons from sources at these positions were moderated by 
part of the inner shielding before reaching the detectors. 

A $^{252}$Cf input spectrum was used
to simulate incident neutron energies, 
and this spectrum was degraded in energy by propagation through 
the inner shielding. 
Features in the input spectrum are washed out to the extent that 
an independent simulation with a Maxwellian input spectrum 
produced an identical recoil energy
spectrum in the detectors, to within statistical uncertainties. 
It is therefore inferred that the spectrum of recoil 
energies for this configuration is largely independent of details of the input 
neutron energy spectrum and is thus sufficiently accurate. 
See \ref{app:A-2} for additional details.
The angular dependence of the differential neutron-scattering cross section for silicon in 
\geant is known to be incorrect~\cite{RobinsonSimLibrary}, 
but using the correct dipole anisotropy moment 
produces an identical nuclear recoil spectrum for 
neutrons scattering in silicon.

These simulated neutron calibration data sets were used to produce the expected energy 
spectra for nuclear recoil events for each detector and source position.
The spectra from measured calibration data were then compared to these expected spectra.
In the simulation, an event's recoil energy in each detector was determined by directly 
summing the energy depositions to recoiling nuclei. 

\subsection{Data Selection Cuts and Efficiency Corrections}

\begin{figure}
\centering
\includegraphics[width=0.55\linewidth]{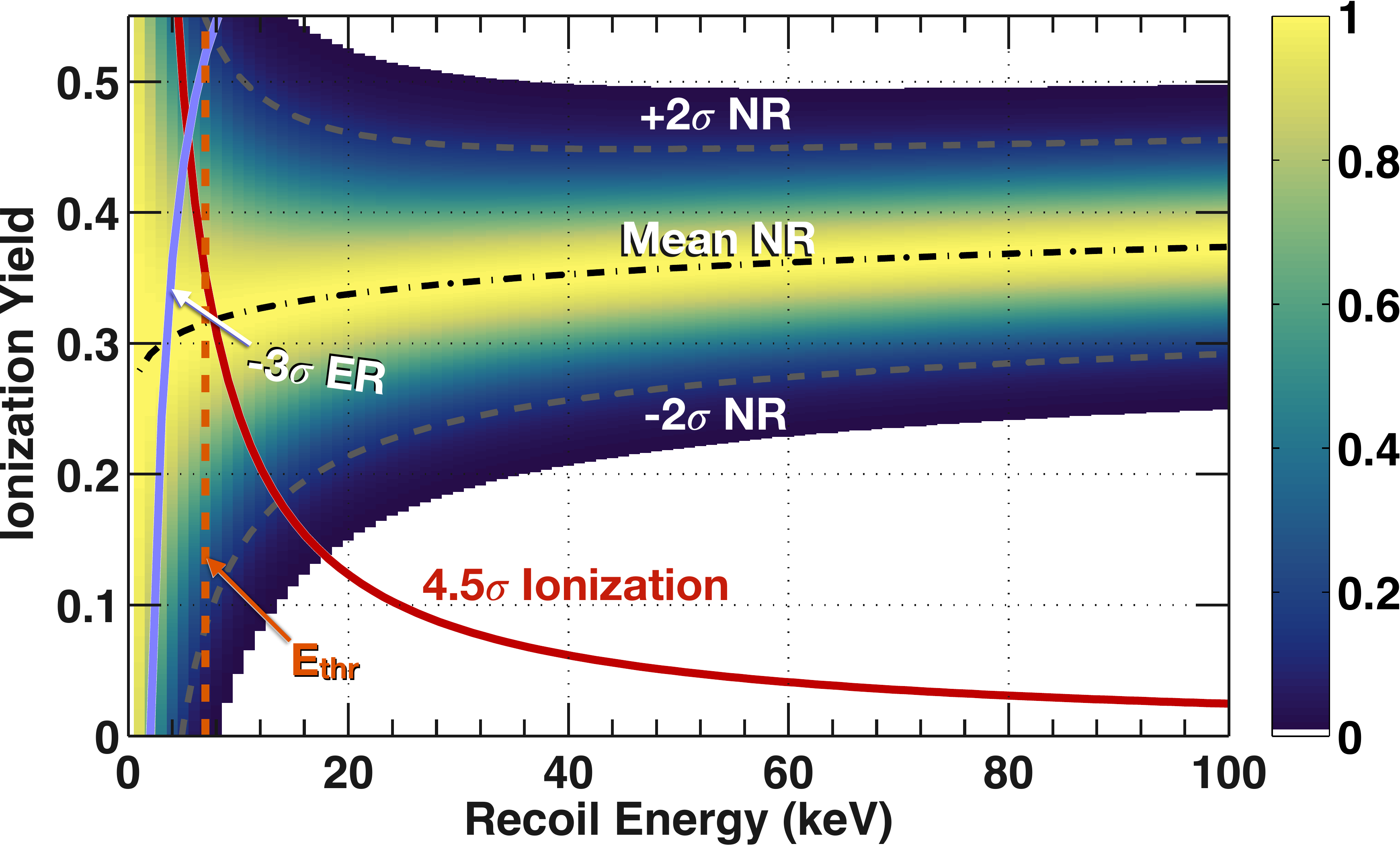}
\caption{Two-dimensional histogram of the probability distribution function 
of nuclear recoils for T1Z4. Bins with less than 1\% are white. 
Events selected as WIMPs must lie within the $\pm2\sigma$ contours (dashed gray) 
surrounding the nuclear-recoil mean (dot-dash black).
Selected events must also lie above the ionization threshold (solid dark red), 
to the right of the analysis threshold (at 7\,keV for this run; dashed orange),
and below the lower $3\sigma$ bound of the electron-recoil band (solid light purple).}
\label{fig:nuclearRecoilBands}
\end{figure}

A sample of good recoil events was selected from the measured calibration data, 
as described in Sec.~\ref{sec:electronRecoilCalib}, 
with the addition of requiring that events fall within $\pm2\sigma$ of the mean measured nuclear-recoil ionization yield 
(as shown in Fig.~\ref{fig:nuclearRecoilBands}).
To correspond to the energy range analyzed in  Ref.~\cite{si_CDMSII_PRL}, the reconstructed 
recoil energy of each event was restricted to lie below 100\,keV 
and above a detector- and run-dependent threshold ranging from 7 to 15\,keV, 
determined primarily by the ionization threshold of the detector for the run (also shown in Fig.~\ref{fig:nuclearRecoilBands}).

\begin{figure}
\centering
\includegraphics[width=0.55\linewidth]{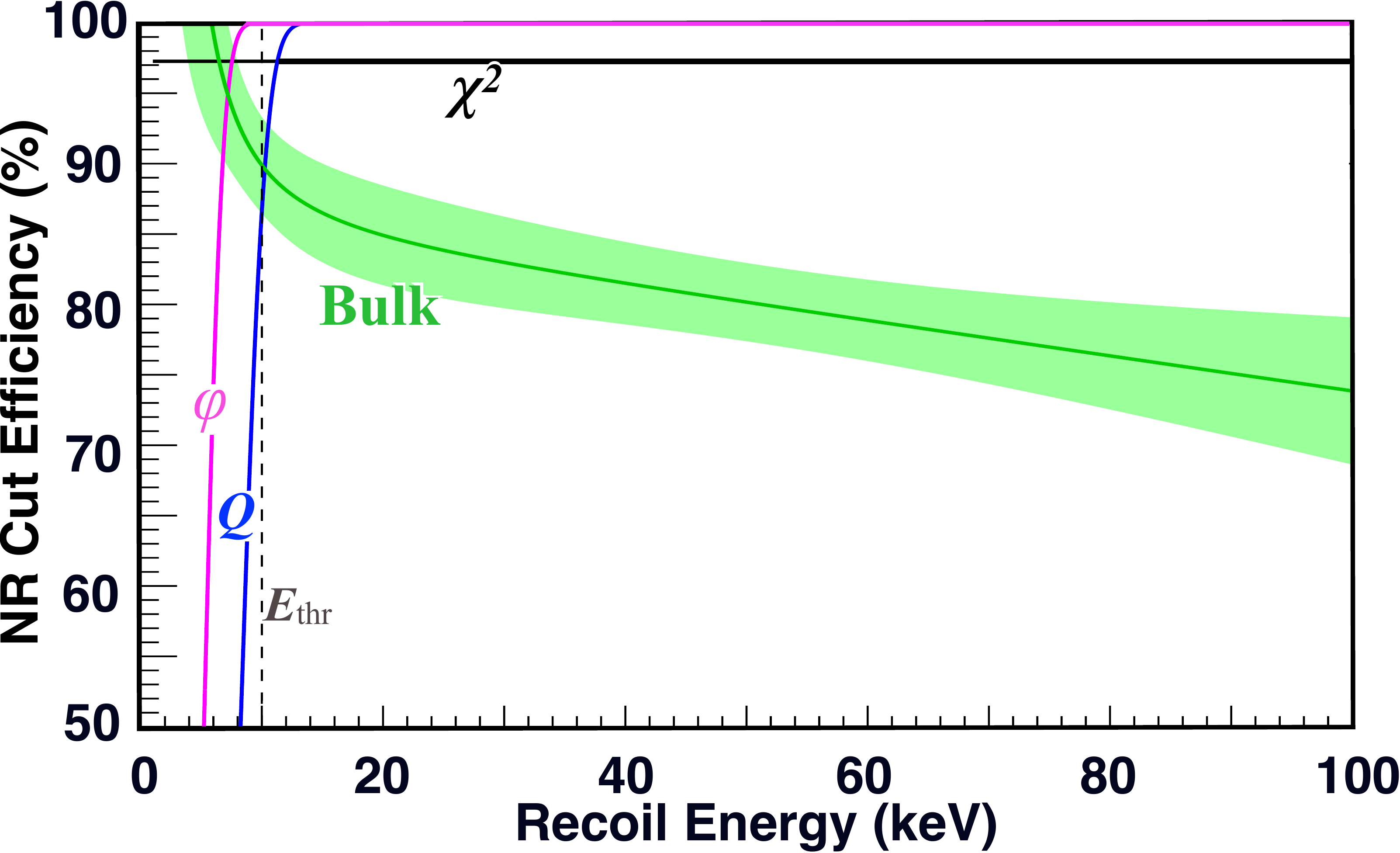}  
\caption{Cut efficiencies as a function of recoil energy for T1Z4. 
The phonon trigger efficiency ($\varphi$, magenta) is unity 
above the analysis energy threshold~$E_\textrm{thr}$ (dashed vertical line at 10~keV). 
The ionization threshold efficiency ($Q$, blue) dominates the determination of the analysis threshold. 
The ionization pulse shape $\chi^2$ efficiency ($\chi^2$, black) has negligible energy dependence. 
The fiducial-volume cut efficiency (Bulk, green) is shown with shaded 1$\sigma$ uncertainty band.}
\label{fig:si_nr_efficiencies}
\end{figure}

There are four potentially important energy-dependent efficiencies in this analysis, 
the forms of which are shown for one detector in Fig.~\ref{fig:si_nr_efficiencies}.
The first is the efficiency of the hardware phonon trigger which 
is modeled as an error function, with a width 
determined by the resolution of the pulse measurement.
The second is the efficiency of the ionization-threshold cut, 
which is the primary determinant of the overall analysis threshold.
The cut's main purpose is to remove sidewall surface events,
which can result in 
no detected ionization~\cite{c38lowmass}. 
Its efficiency is calculated analytically for a 
given recoil energy 
by finding the integrated fraction of the Gaussian probability distribution 
(as shown in Fig.~\ref{fig:nuclearRecoilBands}) that remains within the bounds of the measured 
2$\sigma$ nuclear-recoil band after removing the portion of the band that falls below the ionization threshold.
The efficiency of this cut is estimated in combination with that of 
the cut requiring events to have ionization yield at least 3$\sigma$ below 
the mean of the electron-recoil band. The latter cut ensures that the sample of 
nuclear recoils is not significantly contaminated by electron recoils. 

The remaining efficiencies are those of the $\chi^{2}$ 
goodness-of-fit and fiducial-volume cuts.  
The former rejects poorly shaped ionization pulses 
and has negligible energy dependence~\cite{Jeff_thesis},  
and the latter excludes events occurring in the outer 
edge of the detector where incomplete ionization collection 
can cause electron recoils to mimic nuclear recoils~\cite{c38lowmass}.
The fiducial-volume cut efficiency is calculated for events 
in the nuclear-recoil band, 
including a correction based on an estimate of the
number of electron recoils that leak 
into the nuclear-recoil band~\cite{kmccarthythesis}. 
This efficiency has the strongest energy dependence of all the cuts.

\begin{figure}[h] 
\centering
\includegraphics[width=0.65\linewidth]{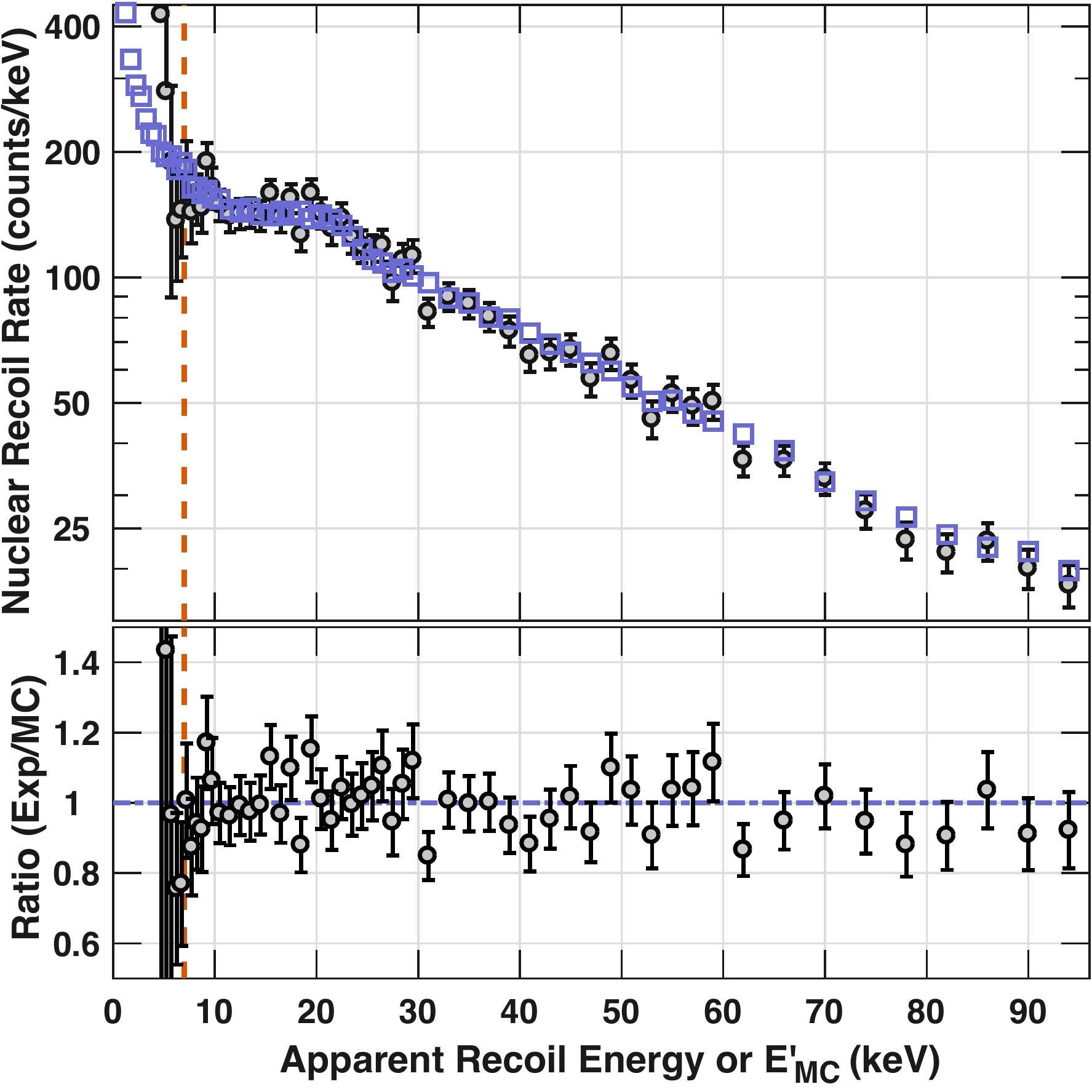}
\caption{{\bf Top:} (Log-scale) Efficiency-corrected nuclear-recoil energy spectrum 
for detector T2Z1 ($\circ$ with 1$\sigma$ error bars)
as a function of the apparent recoil energy
(assuming phonon collection efficiency $\eta^\varphi_{\rm NR}=1$), 
compared with the corresponding simulated spectrum  
after application of the overall best-fit energy scale: 
$E^{'}_{\rm MC} = 0.95 E^{\rm}_{\rm MC}$  
($\square$~with uncertainties smaller than the marker). 
The detector's analysis energy threshold (vertical dashed line) is 7\,keV for this run. 
{\bf Bottom:} Ratio of the efficiency-corrected spectrum to the (overall best-fit) rescaled simulated spectrum.
\label{fig:recoil_spectrum_with_residual}}
\end{figure}

After applying these event selection criteria and efficiency corrections, 
and accounting for the detector masses, 
the resulting spectra give the raw nuclear recoil rate in counts \si{\per\keV\per\kg\per\day}, 
and as such are directly comparable to the spectra  generated by the Monte 
Carlo simulation. Figure~\ref{fig:recoil_spectrum_with_residual} shows an example.

\subsection{Determining the Relative Phonon Collection Efficiency}

The nuclear-recoil energy spectrum in CDMS~II silicon detectors 
is characterized by a single smooth exponential in the energy range 
of interest with a prominent feature 
at $\sim$20\,keV 
caused by a wide nuclear resonance with 
incident neutrons of $E\approx183$\,keV,
as discussed in \ref{app:A}. 
This feature (shown in Fig.~\ref{fig:recoil_spectrum_with_residual}) 
breaks the degeneracy between the rate normalization and 
spectral hardness, making it possible 
to infer the phonon collection efficiency $\pce$ of nuclear recoils relative to that of electron recoils
by comparing measured and simulated spectra  
without knowing the rate of nuclear recoils. 

\begin{figure}[t]
\centering
\includegraphics[width=0.65\linewidth]{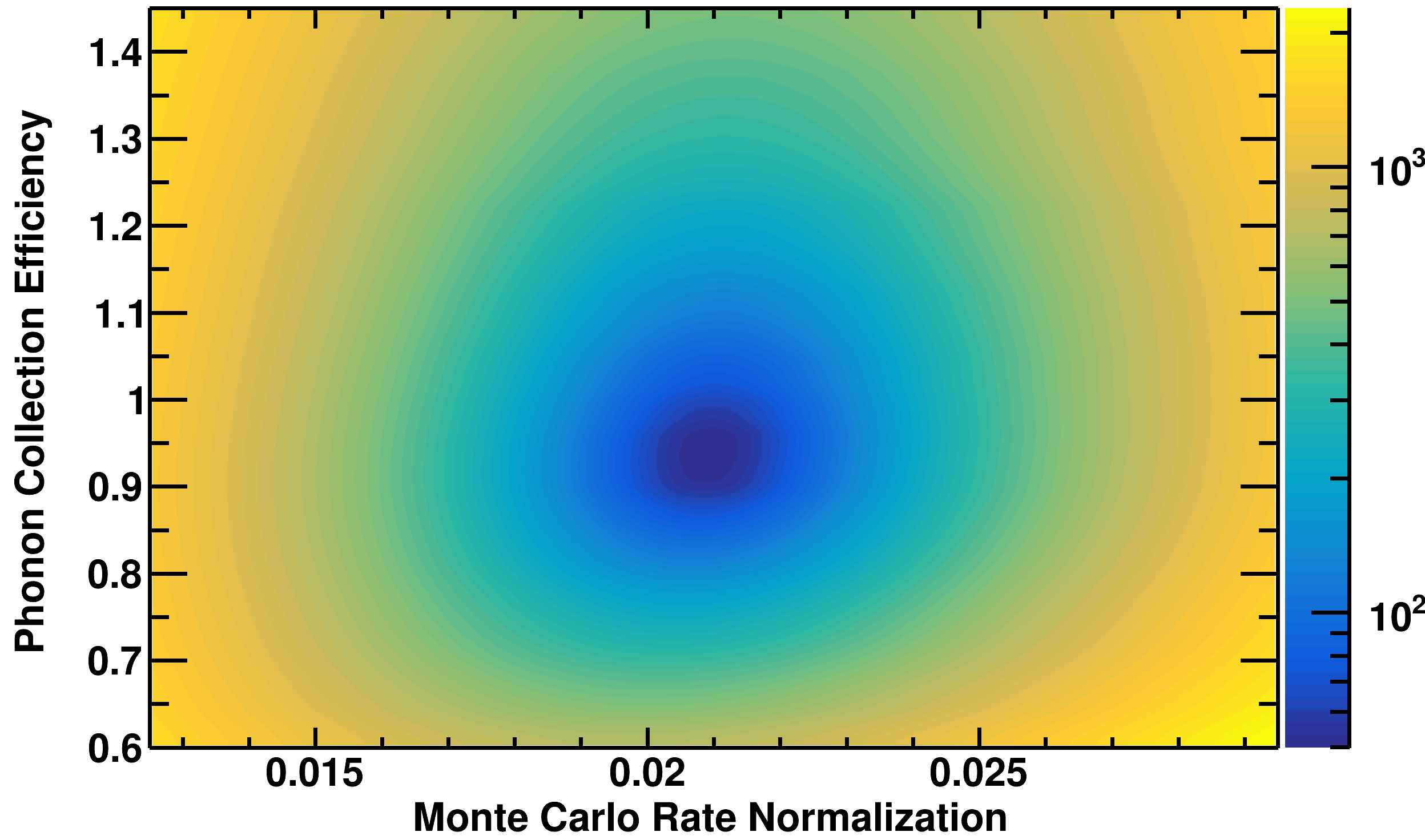}
\caption{(Log-scale) Two-dimensional $\chi^2$ contour map
from Eqn.~\ref{eqn:chi2} for T2Z1---calculated using simulated and measured data coadded over all source positions---as a function of the phonon collection efficiency 
$\pce$ applied to the simulated nuclear recoil energies 
and the Monte Carlo rate normalization. 
The simulated nuclear recoil energies are rescaled to 
$\eta^\varphi_{\rm NR} E^{\rm}_{\rm MC}$. 
\label{fig:MC_chi2contour}} 
\end{figure}

A test value for this energy rescaling factor $\pce$ is applied to the recoil 
energy of each event in the simulated data set prior to binning 
(as in, \textit{e.g.}, Fig.~\ref{fig:recoil_spectrum_with_residual}).
A $\chi^2$ statistic is then constructed from each pair of binned spectra in a way that
incorporates the Poisson errors for each energy bin $i$ of both the measured $(X)$ and 
simulated $(\mu)$ rates: 
\begin{eqnarray}\label{eqn:chi2}
\chi^2 = \sum\limits_ i\left(\frac{X_i-\mu_i}{\sigma_i}\right)^2,
\end{eqnarray}
with $\sigma_i^2 = \sigma_{\mathrm{exp},i}^2 + \sigma_{\mathrm{MC},i}^2$ 
in terms of the measured ($\sigma_{\mathrm{exp},i}$) 
and simulated ($\sigma_{\mathrm{MC},i}$) Poisson uncertainties. 
The energy rescaling is applied to the simulated, 
rather than the measured, data to avoid problems associated 
with event energies shifted above and below threshold 
and to simplify the accounting of energy-dependent efficiencies. 
The simulated nuclear-recoil energies 
are rescaled to $E^{'}_{\rm MC} = E^{\rm}_{\rm MC} \pce$ 
(as in Fig.~\ref{fig:recoil_spectrum_with_residual}).
For each combination of detector and source position, 
a two-dimensional $\chi^{2}$ minimization was performed, 
scanning over both $\pce$ and the normalization.
An overall best-fit $\pce$ for each detector 
was also determined by performing an additional scan, 
after coadding data from all three source positions 
for both the measured and simulated spectra.
This overall rescaling $\pce$ is used to scale 
the apparent nuclear recoil energies by 1/$\pce$. 
Figure~\ref{fig:MC_chi2contour} shows the results of
this minimization for a representative detector.

\begin{figure}[!t]
\centering
\includegraphics[width=0.6\linewidth]{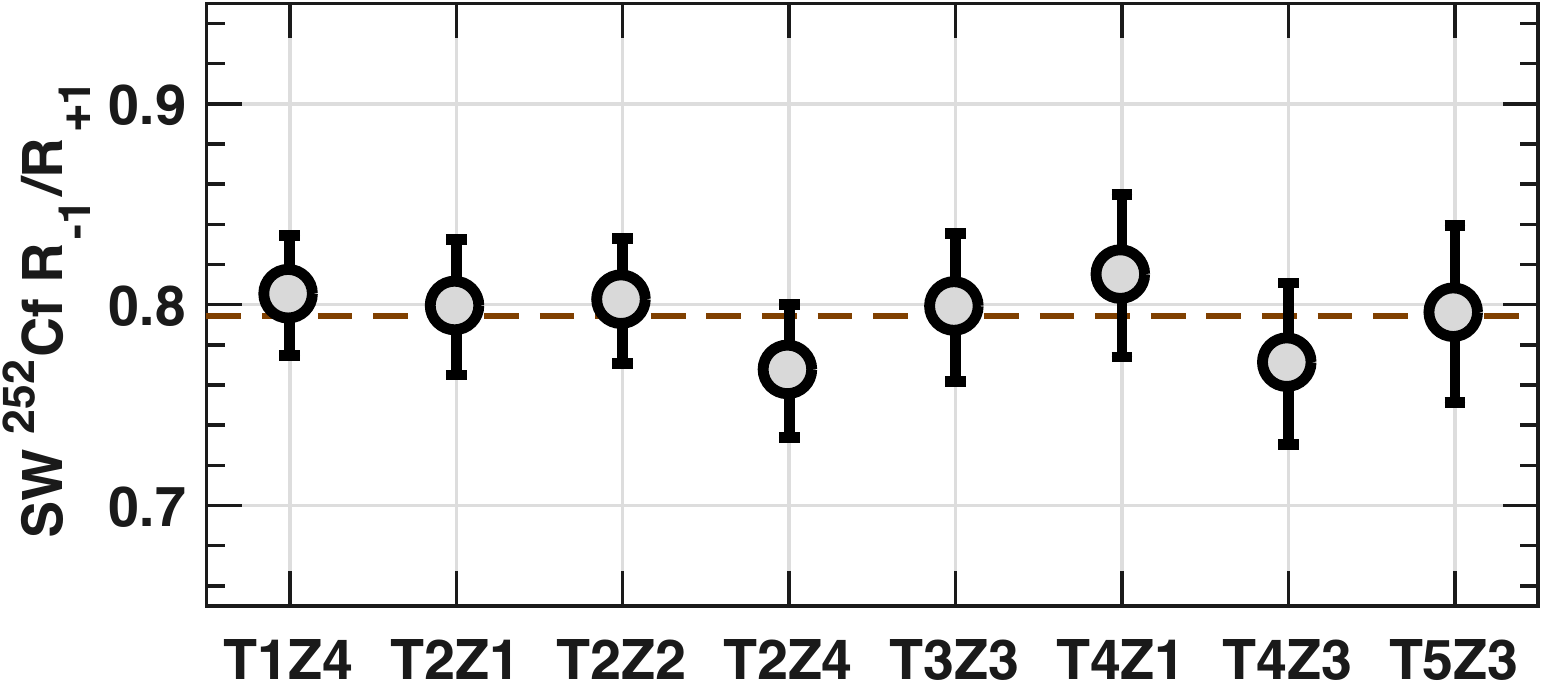}
\caption{Ratio of simulated nuclear recoil rates $R_{-1}/R_{+1}$ 
for the SW source position with the location of the source 
varied by $\pm$1~cm from the nominal source location with Poisson uncertainties.  
The relative detector rates are consistent to within 
statistical uncertainties, but the overall rate (dashed line) decreases by $\sim$\,20\%.
\label{fig:neutron_rate_variation}}
\end{figure}

The overall neutron rate is not used as a constraint 
because it is not known sufficiently well, 
primarily due to uncertainty in the placement of the $^{252}$Cf source
between each calibration.  
As shown in Fig.~\ref{fig:neutron_rate_variation}, 
variation in source placement of $\pm1$\,cm 
changes the rate in all detectors by approximately $\pm$10\%.
The placement of the source was done 
with no way to verify its location with more precision than a centimeter.
The absolute rates measured by the detectors varied
by as much as a factor of~3, even between good $^{252}$Cf calibrations (those for which the detectors were operating properly); 
however, the relative rates for good calibrations were consistent. 
Because most detectors did not record good data throughout the entire exposure, 
care was taken in forming the overall normalization 
to account properly for periods of lost live time in detectors.
This was done by compensating for lost live time during bad series
by weighting the lost live time
by the relative neutron rate inferred from periods of good neutron exposure
from all detectors,
using an iterative fitting procedure.

The best-fit relative normalizations from the $\chi^2$ 
minimization procedure agree at the 90\% confidence level 
for data at two of the source 
positions, with slightly worse agreement at the third position. 
Moreover, as shown in Fig.~\ref{fig:MC_chi2contour},
the best-fit energy rescaling factor is independent of the normalization, 
due to the feature in the spectrum; so the accuracy of the normalization is not important to the results.
The good agreement between the data and rescaled Monte Carlo 
across the entire energy region of interest indicates that 
the overall best-fit phonon collection efficiency is close to 100\%.

\section{Results}
\begin{figure}[pt!]
\centering
\includegraphics[width=0.7\linewidth]{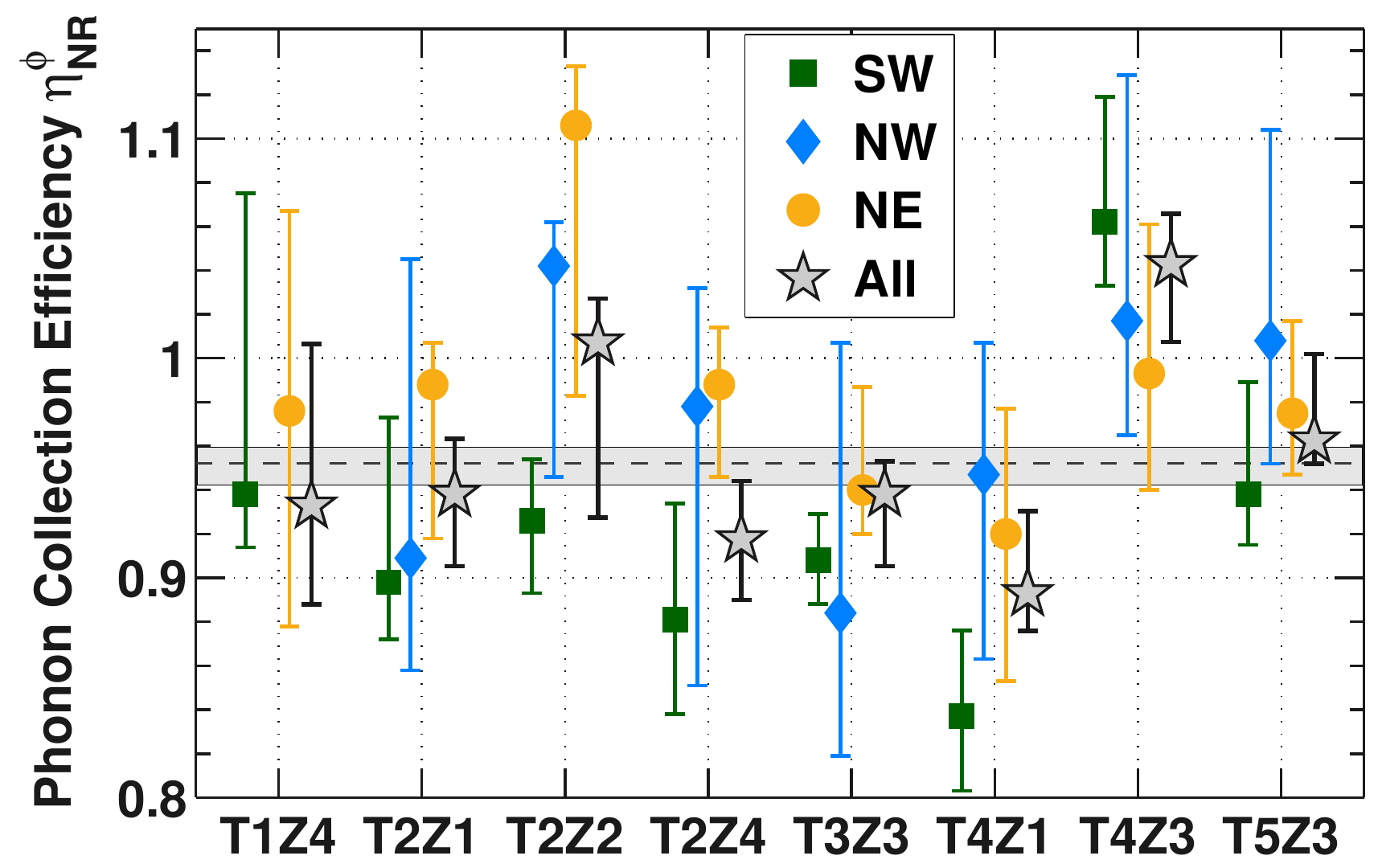} 
\caption{Best-fit phonon collection efficiency for 
SW~($\blacksquare$), NW~($\blacklozenge$), NE~($\bullet$),  
and coadded~($\bigstar$) $^{252}$Cf source positions
for each detector at 95\% C.L. 
Most detectors (except for T2Z2 and T4Z3) 
show an underestimation of nuclear recoil energy. 
The weighted average over all detectors of the 
coadded best-fit results (gray fill region) 
gives $\pce = 95.2^{+0.9}_{-0.7}$\,\% at 95\% C.L. 
No acceptable NW neutron calibration data sets exist for detector T1Z4.
\label{fig:bestfit_recoil_scale}}
\end{figure}

The final result of the $\chi^2$ minimization is a best-fit 
phonon collection efficiency $\pce$ for nuclear recoils 
relative to electron recoils for each detector and source position, 
shown in Fig.~\ref{fig:bestfit_recoil_scale}. 
A weighted average across all silicon detectors, 
using the best-fit results from the individual-detector fits 
(coadded over source position), 
finds an overall phonon collection efficiency for nuclear recoils
\begin{eqnarray*}
\label{eqn:bfpce}
\pce = 95.2^{+0.9}_{-0.7}\%
\end{eqnarray*}
relative to electron recoils of the same deposited energy. 
Table~\ref{tab:chi2_pval} lists the best-fit $\chi^{2}$/d.o.f.\ and $p$-value for each detector. 
The discrepancies
in the best-fit $\pce$ between detectors cannot be explained by 
energy dependence in the relative collection efficiency $\pce$, 
because the measured and simulated spectral data generally match well
both at low energies near the prominent 20~keV feature, 
and at energies up to 100~keV, as shown in the bottom of 
Fig.~\ref{fig:recoil_spectrum_with_residual}.
\begin{table}[t]
\centering
\resizebox{\columnwidth}{!}{
\begin{tabular}{c c c c c c c c c}
\hline
detector & T1Z4 &  T2Z1 & T2Z2 & T2Z4 & T3Z3 & T4Z1 & T4Z3 & T5Z3 \\
\hline
$\chi^2$/d.o.f. & 40.1/42 & 50.0/48 & 72.7/48 & 75.0/48 & 57.5/48 & 53.3/42 & 33.6/37 & 81.6/48\\
$p$-value & .554 & .394 & .012 & .008 & .164 & .113 & .629 & .002 \\  
\hline
\end{tabular}
}
\caption{Minimum $\chi^2$/d.o.f.~by detector, 
using coadded spectral data from all source positions,  
of the best-fit phonon collection efficiency $\pce$ 
for nuclear recoils relative to electron recoils.
Detectors with higher energy thresholds have fewer energy 
bins and therefore fewer degrees of freedom.}
\label{tab:chi2_pval}
\end{table}

\subsection{Implications for Ionization Yield}
 
The stopping power for charged particles in a target material can 
be divided into electronic and nuclear components, 
each with different energy dependence as reported in Ref.~\cite{lindhard}. 
Slow-moving nuclear recoils are not stopped efficiently 
by electrons and so deposit most of their energy through 
interactions with the target's nuclei. 
Because ionization is a product of electronic excitation, 
nuclear recoils have a reduced yield compared to electron recoils of the same energy.
The ionization yield of a nuclear recoil varies with the partitioning of energy between electronic and nuclear modes. 
The energy dependence of the reduced yield as a function of atomic number $Z$ and atomic mass $A$ was computed by Lindhard in Ref.~\cite{lindhard}. 
The resulting expressions were simplified and reported in Ref.~\cite{lewinsmith}. 
The expected ionization yield for a nuclear recoil 
under this Lindhard theory
is given by
\begin{eqnarray}
\label{eqn:lindhard}
y_L = \frac{k\,g(\varepsilon)}{1+k\,g(\varepsilon)}, 
\end{eqnarray}
where $k= 0.133\,Z^{2/3} A^{-1/2} \approx 0.146$ for silicon, 
and the transformed energy $\varepsilon = 11.5\,E_{\rm R}\,Z^{-7/3}$,
with the recoil energy $E_{\rm R}$ given in keV. 
The function $g(\varepsilon)$ is well-fit by a polynomial in $\varepsilon$ with empirically chosen coefficients, described by $3\varepsilon^{0.15} + 0.7\varepsilon^{0.6} +\varepsilon$. 

\begin{figure}
\centering
\includegraphics[width=0.75\linewidth]{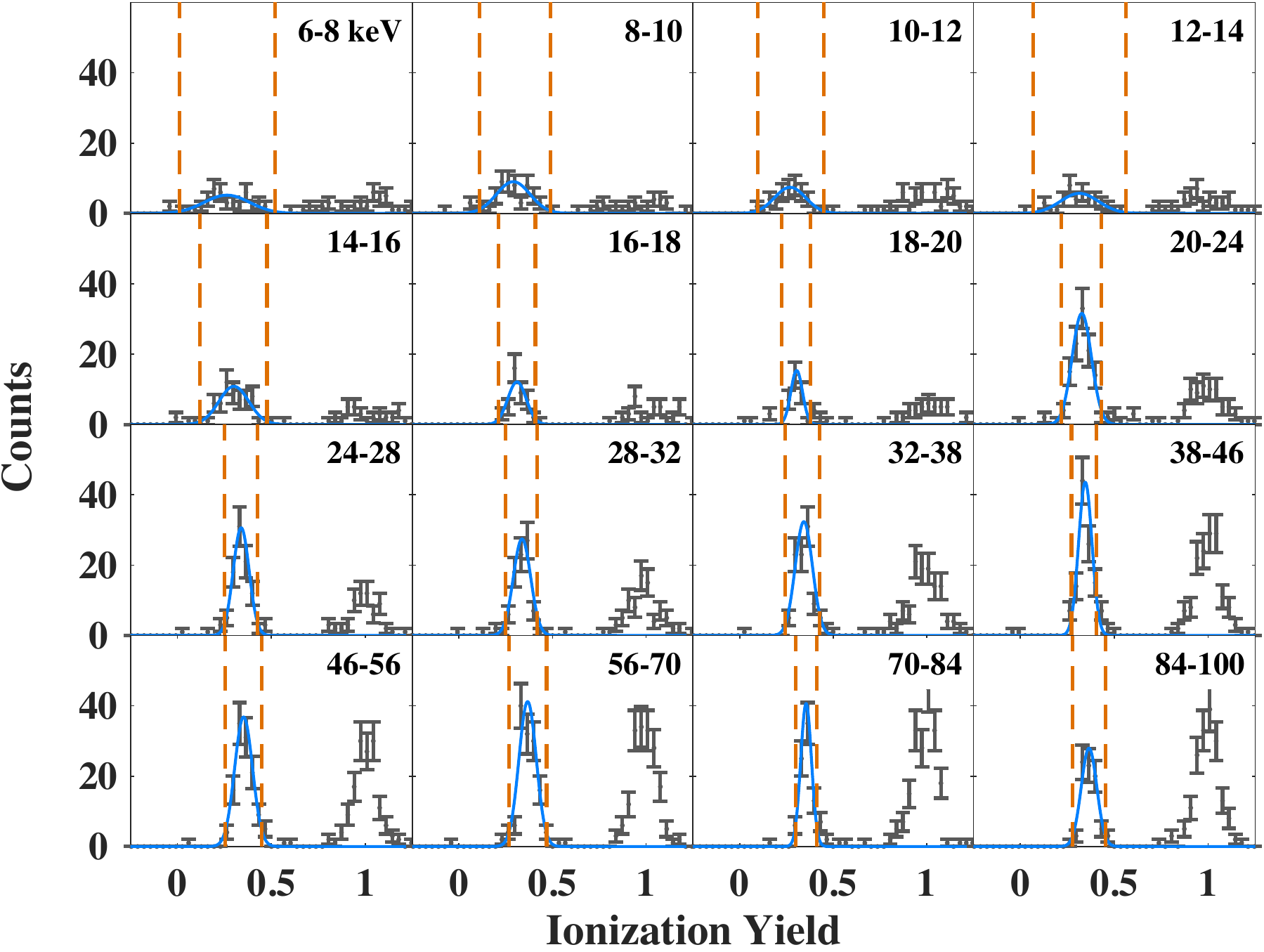} 
\caption{Fits to the nuclear-recoil ionization yield for detector T1Z4 $^{252}$Cf data, 
performed in bins of recoil energy (rescaled by $\pce=0.952$) from 6--100\,keV. 
In each bin, a Gaussian distribution (solid line) is fit to the observed counts 
(with 1$\sigma$ error bars) within the indicated ionization-yield range (vertical dashed) for nuclear recoils.}
\label{fig:full_fit}
\end{figure}

The same neutron calibration data discussed 
in Sec.~\ref{sec:NRcalibration} were used to infer the  
ionization yield of nuclear recoils in CDMS~II silicon detectors.
Figure~\ref{fig:full_fit} shows fits to the 
measured ionization yield in bins of recoil energy, 
corrected by the best-fit phonon collection efficiency ($\pce$ = 0.952)
for an example detector.
The resulting inferred ionization yield as a function of 
recoil energy must be corrected for the small effect of neutron multiple scattering. 
While WIMPs have a negligible probability of 
scattering more than once in the apparatus, 
approximately 30\% of neutrons from 2--100\,keV 
scatter off nuclei at multiple locations in a single detector. 
The ionization yield of nuclear recoils increases with increasing recoil energy. 
Hence a multiple-site interaction, 
for which the ionization is divided among several lower-energy recoils, 
will produce less ionization (overall) 
than a single recoil of the same total recoil energy.
These multiple-site scatters are not distinguishable 
from single-site interactions of the same total energy
in the CDMS setup.
The effect of multiple scattering was determined from 
\geant simulations of the $^{252}$Cf neutron calibrations.
The shifts in yield are well-understood and 
are less than 3\% 
for nuclear recoils between 10 and 100\,keV.  

\begin{figure}
\centering
\includegraphics[width=0.7\linewidth]{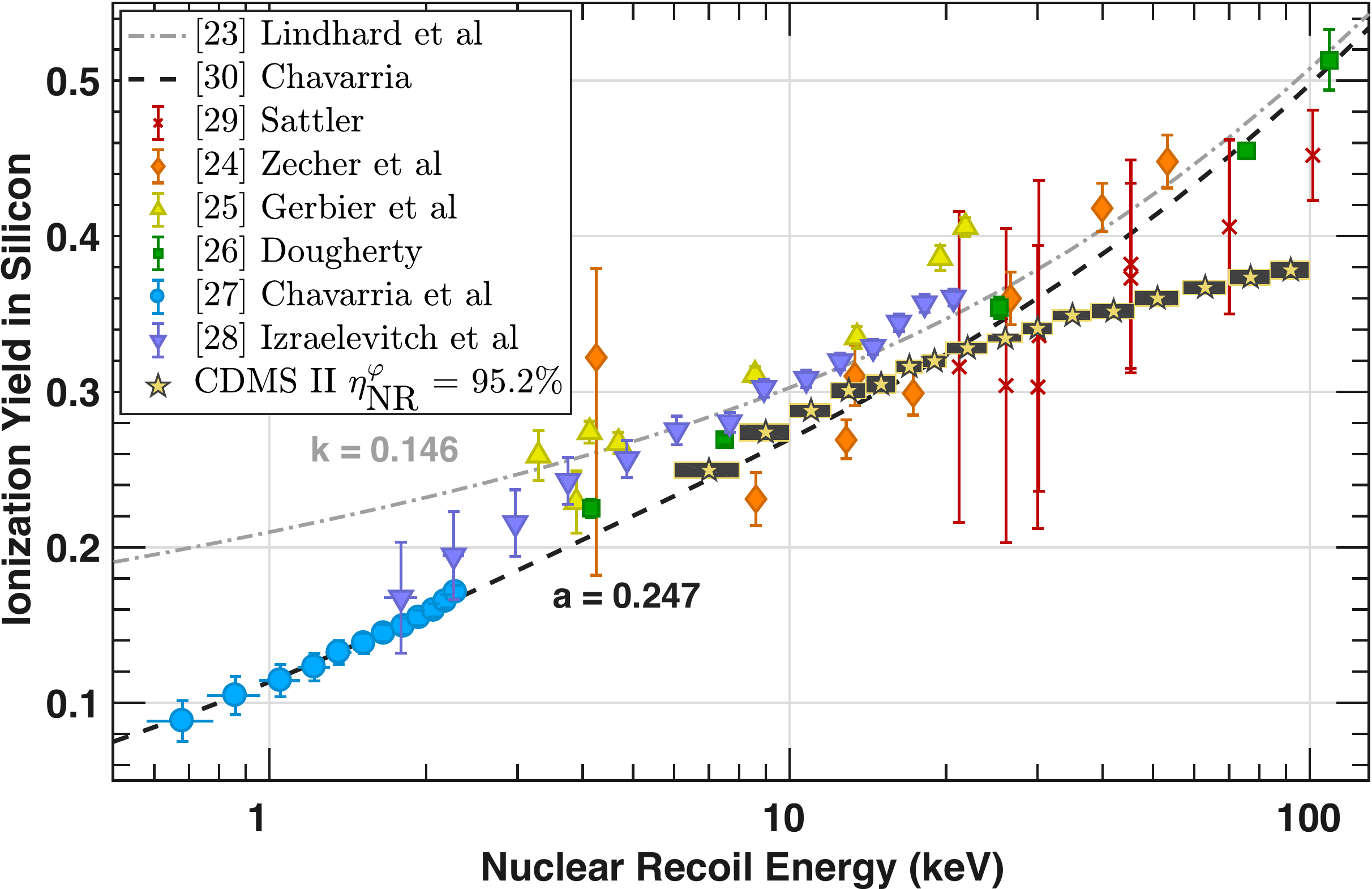}
\caption{Measurements of the ionization yield for nuclear recoils 
in silicon~\cite{PhysRevA4058,PhysRevD3211, PhysRevA2104, damic2016ionization, antonella2017_lowE_NRs, PhysRevA1815}. 
The light dot-dashed line shows the 
theoretical prediction $y_L$ for silicon ($k=0.146$) from Lindhard~\cite{lindhard}.
The dark dashed line shows the
parameterization $y_C$ (with $a=0.247$) 
from Chavarria~\cite{chavPrivateComm},
which fits the existing data reasonably well. 
Data from this analysis ($\bigstar$) are the 
weighted means of ionization yield for the eight silicon detectors 
including phonon energy rescaling and multiple-scattering corrections, 
with uncertainty bands representing the standard deviation
and the nuclear-recoil energy-bin width.}
\label{fig:LitComp_CDMS}
\end{figure}

In silicon, the yield has been measured previously by 
elastic scattering at 77\,K~\cite{PhysRevA4058,PhysRevD3211,PhysRevA2104}, 
130\,K~\cite{damic2016ionization}, 
220\,K~\cite{antonella2017_lowE_NRs}, and 
288\,K~\cite{PhysRevA1815},
and by utilizing resonances in the scattering cross section
to constrain the recoil energy~\cite{PhysRevA2104}. 
The results of these previous measurements are 
summarized in Fig.~\ref{fig:LitComp_CDMS}. 

The light dot-dashed line shows the standard 
Lindhard theoretical prediction $y_L$ (from Eq.~\ref{eqn:lindhard}) 
for ionization yield in silicon from Ref.~\cite{lindhard}.
Standard Lindhard theory significantly over-estimates
the ionization production for low-energy nuclear recoils 
reported in Ref.~\cite{damic2016ionization}.  
An improved functional form (black dashed) using a parameter $a=0.247$~\cite{chavPrivateComm} 
matches the Lindhard expectation 
$y_L$ for silicon at high energy 
and fits the data reported in Ref.~\cite{damic2016ionization}
at low energy: 
\begin{eqnarray}
\label{eqn:chavParm}
y_C = \left(\frac{1}{a E_{\rm R}} + \frac{1}{y_L}\right)^{-1}.
\end{eqnarray}
This parameterization was used to report the WIMP-nucleon sensitivity
curves in Ref.~\cite{antonella2017_lowE_NRs}. 
CDMS~II silicon data are consistent 
with this functional form for energies below 20\,keV.
At high energies, the measured ionization yield 
is smaller than previous measurements 
\cite{PhysRevA4058, PhysRevA2104, PhysRevA1815},
with the size of the discrepancy increasing with energy.

\begin{figure}
\centering
\includegraphics[width=0.65\linewidth]{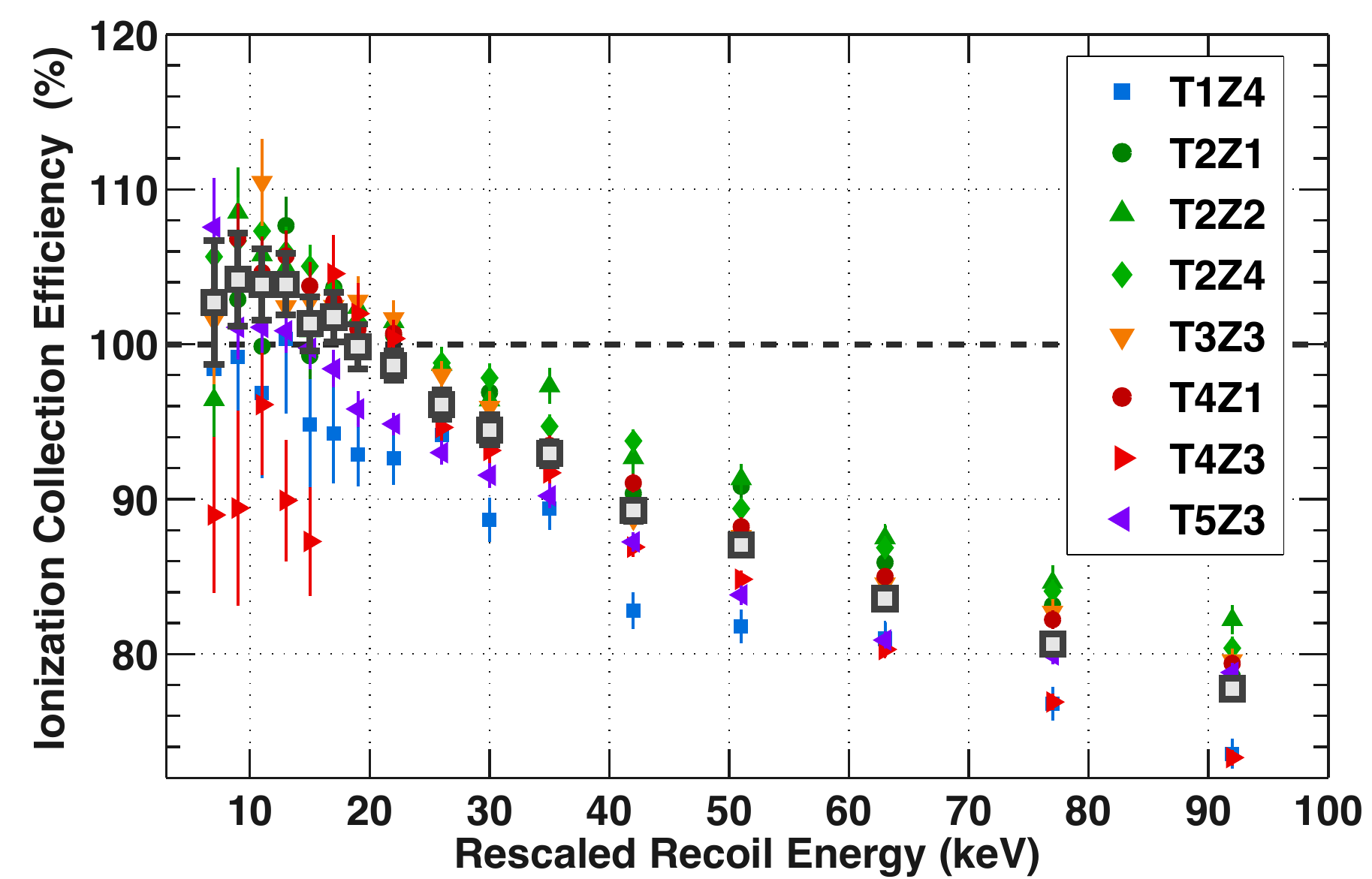} %
\caption{Energy-rescaled measurements of the (multiple-scatter corrected)  
ionization collection efficiency {\it vs.}\ recoil energy for nuclear recoils in CDMS~II silicon detectors. 
The ionization collection efficiency assumes the parameterization for $y_C$ from Ref.~\cite{chavPrivateComm}. 
The error bars indicate the results of the fits 
to the ionization yield distributions for each detector individually,
and weighted means (bold $\square$) with $1\sigma$ error bars
$\sigma_\mu = \sigma_{\rm y} / \sqrt{8}$
where $\sigma_{\rm y}$ is the standard deviation of the 
8 best-fit detector yields for each energy bin. 
The results are consistent with 100\% ionization collection efficiency (gray dashed) 
at energies <20~keV but gradually decrease to $\sim$75\% at 100~keV. 
}
\label{fig:cdms_meas}
\end{figure}
Figure~\ref{fig:LitComp_CDMS} shows the 
ionization yield determined from the Gaussian fits 
to the nuclear-recoil distribution for each detector 
(as in Fig.~\ref{fig:full_fit}, \textit{e.g.}). 
Comparisons of the CDMS~II measurements of the yield to 
the previous measurements 
shown in Fig.~\ref{fig:LitComp_CDMS} constrain the 
nuclear-recoil ionization collection efficiency in CDMS~II. 
Figure~\ref{fig:cdms_meas} shows the 
nuclear-recoil ionization collection efficiency 
for all CDMS~II silicon detectors and their weighted mean, 
assuming the same parameterization~\cite{chavPrivateComm}
shown in Fig.~\ref{fig:LitComp_CDMS}. 
The individual detector fits are not consistent 
with each other within uncertainties.
These detector-to-detector variations 
may correspond to true physical differences between the detectors.
The average ionization collection efficiency for nuclear recoils
in CDMS~II silicon ZIPs 
is consistent with roughly 100\% at energies below 20\,keV.
The fit is improved if the ionization-yield parameterization
from Eq.~\ref{eqn:chavParm} underestimates the true ionization yield by $\sim$5\% from 10--20\,keV.

There is a reasonable mechanism for producing 
the ionization collection efficiency observed 
in CDMS~II silicon detectors.
The detectors were operated with fields of a few V/cm, 
much lower than those described in 
Refs.~\cite{PhysRevA4058, PhysRevD3211, PhysRevA2104, damic2016ionization, 
antonella2017_lowE_NRs, PhysRevA1815}. 
The electric fields reported in Refs.~\cite{PhysRevA4058} and \cite{PhysRevA2104}  
are 60--500$\times$ larger than the 4~V/cm electric fields applied to the CDMS~II ZIP detectors.
Nuclear recoils produce a much denser initial composite 
of charge pairs than similar-energy electron recoils.
It is plausible that the ionization produced 
by a low-energy nuclear recoil 
may be fully extracted at low fields, 
but the denser ionization 
produced by higher-energy nuclear recoils 
cannot be extracted completely at these low fields.
The resulting ionization collection efficiency 
may decrease with increasing energy 
because the charges are increasingly self-shielded, 
thus allowing a larger fraction of charge pairs to recombine 
before they can be drifted across the detector. 
It is also possible that the lower temperature 
of the CDMS~II detectors relative to those described in
Refs.~\cite{PhysRevA4058, PhysRevD3211, PhysRevA2104, damic2016ionization,
antonella2017_lowE_NRs, PhysRevA1815} 
plays a role.

\subsection{Recalculated WIMP Limits}

\begin{figure}
\centering
\includegraphics[width=0.7\linewidth]{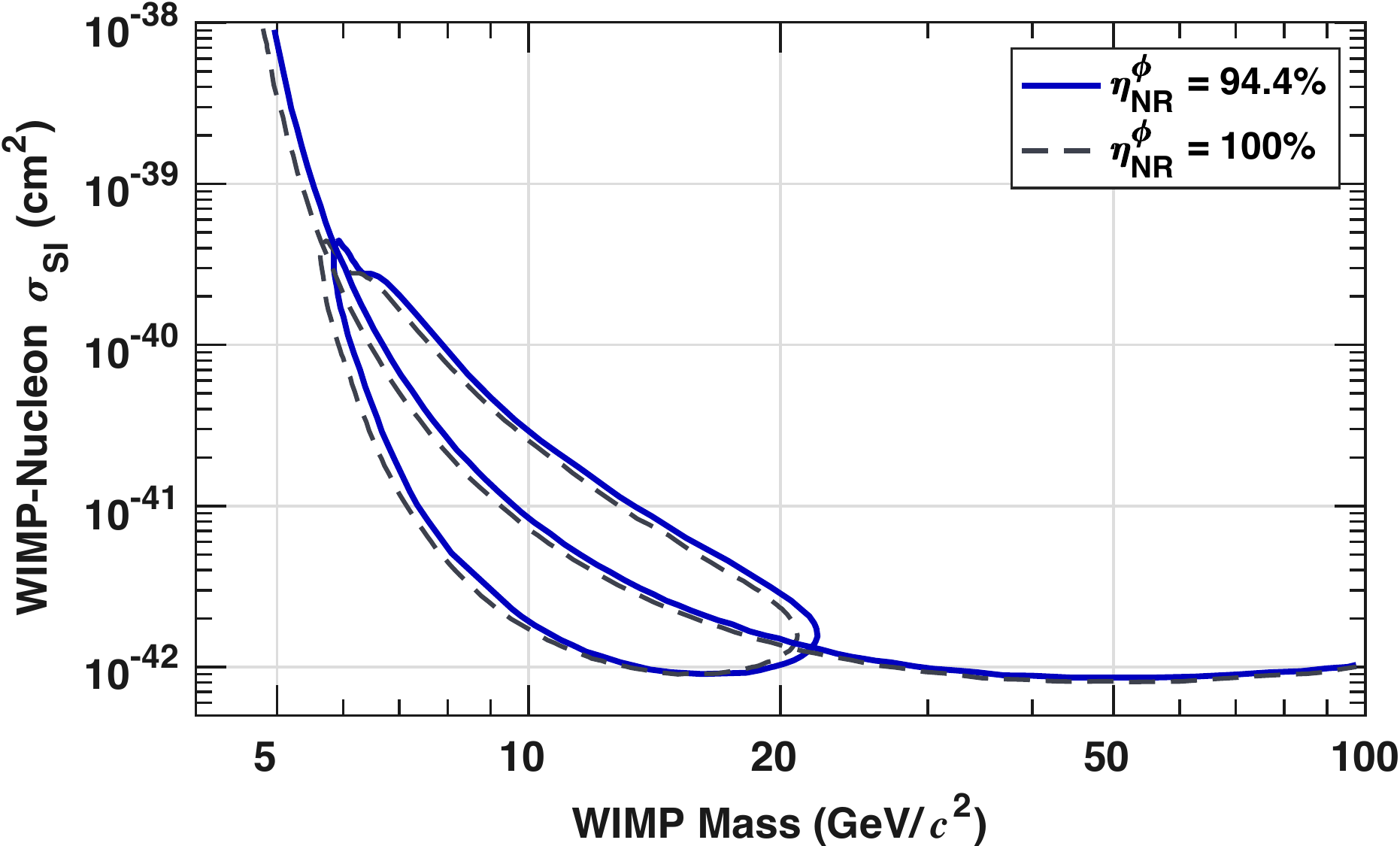}
\caption{90\% C.L. upper limit (curves)  and acceptance contour (closed regions) 
on the spin-independent WIMP-nucleon cross section~$\sigma_\textrm{SI}$ on silicon, 
as published in~\cite{si_CDMSII_PRL} (dashed), 
and using the 1$\sigma$ lower limit 
on the best-fit phonon collection efficiency: $\pce = 94.4\%$ (solid).} 
\label{fig:Si_limit_shift}
\end{figure}

The revised nuclear-recoil energy scale 
has a small effect on published CDMS~II WIMP sensitivity limits and contours.  
Figure~\ref{fig:Si_limit_shift} shows the shifts in 
both the spin-independent WIMP-nucleon cross-section exclusion curve 
and the best-fit WIMP mass region and cross section, both at 90\% C.L., 
from Ref.~\cite{si_CDMSII_PRL}.  
The shifts are generally small. 
For WIMP masses above 10~GeV/$c^2$, 
the shift is less than~20\%, 
and for WIMP masses $\approx$5~GeV/$c^2$
the upper limit increases by about a factor of two.
The best-fit WIMP mass resulting from the revised nuclear-recoil energy scale 
shifts by $<$5\%.

\subsection{Conclusions}
The measured spectral shape of neutron calibration data 
in CDMS~II silicon detectors 
provides strong evidence that the 
phonon collection efficiency $\pce$ 
is almost, but not quite, 
the same for nuclear recoils as electron recoils.
Results are consistent with phonon collection efficiency 
$\pce=95^{+0.9}_{-0.7}\%$ for all energies and detectors, 
with good agreement between measured and simulated spectra down to detector energy thresholds $\sim$10\,keV.
Any energy dependence in $\pce$ in the 10--100\,keV 
energy range considered here cannot be large.
Although similar analysis is possible in germanium, it is prone to systematic uncertainty because the spectrum for this range of nuclear-recoil energies is featureless and decays exponentially.   
As a result, there is an inherent degeneracy between the neutron rate and the energy scale that is difficult to break in the CDMS setup. Imperfect knowledge of the source strength and position make a 
simple comparison of the measured and simulated neutron rates infeasible.
The low-energy nuclear resonance in $^{28}$Si provides a spectral feature that breaks the degeneracy, making the silicon analysis presented here possible.

The CDMS~II silicon ionization measurements described 
in this paper support recent findings of Refs.~\cite{damic2016ionization,antonella2017_lowE_NRs} 
that the Lindhard prediction for nuclear-recoil ionization yield
at low energies ($\lesssim$\,20\,keV) is an over-estimate, 
and that the energy-dependent parameterization of Eq.~\ref{eqn:chavParm} 
is a more accurate description for nuclear recoils in silicon.  
The CDMS data suggest that this parameterization 
may slightly underestimate the true ionization yield
of nuclear recoils between 10 and 20\,keV.
Reduced ionization collection efficiency in CDMS~II 
silicon detectors at recoil energies $\gtrsim$\,20\,keV
may be due to the 
field-dependent self-shielding of charge carriers that prevents them from being drifted by the electric field.
Planned calibration of SuperCDMS SNOLAB~\cite{scdms_PRDSensPaper} silicon detectors
will provide improved measurements of the phonon and 
ionization response, especially at lower energies.

The CDMS Collaboration gratefully acknowledges the 
contributions of numerous engineers and technicians; 
we would like to especially thank Dennis Seitz, Jim Beaty, 
Bruce Hines, Larry Novak, Richard Schmitt, and Astrid Tomada. 
In addition, we gratefully acknowledge assistance 
from the staff of the Soudan Underground Laboratory 
and the Minnesota Department of Natural Resources. 
This work is supported in part by the 
National Science Foundation, 
by the U.S. Department of Energy, 
by the Swiss National Foundation, 
by NSERC Canada, 
and by MULTIDARK. 
This document was prepared by the SuperCDMS collaboration 
using the resources of the Fermi National Accelerator Laboratory (Fermilab), 
a U.S. Department of Energy, Office of Science, HEP User Facility. 
Fermilab is managed by Fermi Research Alliance, LLC (FRA), acting under Contract No. DE-AC02-07CH11359.
Pacific Northwest National Laboratory is operated by Battelle Memorial Institute 
under Contract No. DE-AC05-76RL01830 for the U.S. Department of Energy.
SLAC is operated under Contract No. DE-AC02-76SF00515 
with the U.S. Department of Energy.

\appendix
\section{Numerical Derivation of $^{252}$Cf Spectral Shapes}
\label{app:A}

We review here a numerical calculation of the spectral shapes expected from $^{252}$Cf neutrons scattering from CDMS~II silicon ZIP detectors, adapted from Appendix~E in~\cite{rbunkerthesis}. 
Observation of a prominent bump near 20\,keV in the silicon detectors' $^{252}$Cf spectra in data and simulation prompted the ensuing calculations, both to verify the expectation and to check for other features.  Obviously, any distinguishing features in the $^{252}$Cf spectra are useful for gauging the nuclear-recoil energy scale.  In the following, we derive the precise recoil-energy shapes by using the same {\scshape endf}~\cite{endf} neutron cross sections and angular probabilities that serve as inputs to the \geant simulations.

\subsection{Differential scattering rate}
\label{app:A-1}

The derivation of the differential scattering rate for neutrons scattering from nuclear targets is analogous to the standard framework for WIMP-nucleon scattering (see, {\it e.g.}, \cite{lewinsmith}).  However, the energy dependence is slightly different because the spectrum of incident energies is not defined according to a Maxwellian velocity distribution, but instead is determined (in this case) by transporting the distribution of neutron energies emitted by $^{252}$Cf through the CDMS~II shielding layers.  Treatment of the elastic-scattering cross section differs as well.  In this section we outline a loose derivation aimed toward understanding the energy dependence.  No attempt is made to derive the absolute normalization, with several (constant) multiplicative factors neglected or dropped along the way.

The differential scattering rate for neutrons to scatter from a nuclear target  is given by
\begin{equation}\label{eq:nrap1}
\frac{dR}{dq^{2}} \propto \frac{d\sigma}{dq^{2}}\left(q^{2},v\right)\textrm{\,}v\textrm{\,}n(v),
\end{equation}
where $q^{2}$ is proportional to the transferred energy, $v$ is the relative neutron-nucleus velocity, $n$ is the velocity-dependent neutron number density, and $\sigma$ is the energy- and velocity-dependent neutron-nucleus cross section.  Note that Eq.~\ref{eq:nrap1} is true for a particular value of $v$.  To get the correct recoil-energy shape, the right-hand side of Eq.~\ref{eq:nrap1} must be integrated over all possible velocities:
\begin{equation}\label{eq:nrap2}
\frac{dR}{dq^{2}} \propto \int \frac{d\sigma}{dq^{2}}\frac{dn}{dv}dv = \int \frac{d\sigma}{dq^{2}}\frac{dn}{dE_{\rm i}}dE_{\rm i}\sqrt{E_{\rm i}},
\end{equation}
where the right-hand side is obtained via a change of variables from $v$ to the incident neutron energy $E_{\rm i} \propto v^{2}$.  At this point it is useful to recall the elastic-scattering relationship between the kinetic energy of the recoiling nucleus $E_{\rm R}$, the energy of the incident neutron, and the center-of-mass scattering angle $\theta^{*}$:
\begin{equation}\label{eq:nrap3}
E_{\rm R} = \frac{2A}{\left(1+A\right)^{2}}E_{\rm i}\left(1 - \textrm{cos\,}\theta^{*}\right),
\end{equation}
where $A = 28$ is a good approximation for the atomic mass of a silicon target.\footnote{We consider here only the stable isotopes present in laboratory-grown (non-enriched) silicon crystals with naturally occurring abundances $>$5\%.}  For nonrelativistic scattering, $q^{2} \propto E_{\rm R}$, and the differential cross section can be rewritten as
\begin{equation}\label{eq:nrap4}
\frac{d\sigma}{dq^{2}} \propto \frac{d\sigma}{dE_{\rm R}} = \frac{d\sigma}{d\Omega}\frac{\delta\Omega}{\delta E_{\rm R}} \propto \frac{1}{E_{\rm i}}\frac{d\sigma}{d\Omega},
\end{equation}
because $\Omega \propto \textrm{cos\,}\theta^{*}$ and $\textrm{cos\,}\theta^{*} \propto E_{\rm R}/E_{\rm i}$.  Noting that $dR/dq^{2} \propto dR/dE_{\rm R}$, and substituting the right-hand side of Eq.~\ref{eq:nrap4} into the right-hand side of Eq.~\ref{eq:nrap2}, the differential scattering rate can be written as
\begin{equation}\label{eq:nrap5}
\frac{dR}{dE_{\rm R}} \propto \int \frac{d\sigma}{d\Omega}\frac{dn}{dE_{\rm i}}\frac{dE_{\rm i}}{\sqrt{E_{\rm i}}},
\end{equation}
where the integrand is now entirely in terms of $E_{\rm i}$ and the center-of-mass scattering angle.  For a given value of $E_{\rm R}$, the integral is restricted to combinations of $E_{\rm i}$ and $\textrm{cos\,}\theta^{*}$ that satisfy Eq.~\ref{eq:nrap3}.  Specifically, because $\textrm{cos\,}\theta^{*}$ varies from -1 to 1, the integral runs from $(1+A)^{2}E_{\rm R}/4A$ to $\infty$.  

Equation~\ref{eq:nrap5} and the limits of integration noted above provide the framework necessary to calculate the shape (or energy dependence) of the differential event rate for a spectrum of neutrons to scatter from a nuclear target.  All that remains is to specify the differential number density $dn/dE_{\rm i}$ and the differential cross section $d\sigma/d\Omega$.  The former is simply the spectrum of incident neutron energies, while the latter decomposes into two parts:
\begin{equation}\label{eq:nrap6}
\frac{d\sigma}{d\Omega} \propto \sigma(E_{\rm i})P\left(\textrm{cos\,}\theta^{*} | E_{\rm i} \right),
\end{equation}
where $\sigma(E_{\rm i})$ is the elastic cross section as a function of incident neutron energy (analogous to the WIMP-nucleus form factor), and $P\left(\textrm{cos\,}\theta^{*} | E_{\rm i} \right)$ is the angular probability for a particular value of $\textrm{cos\,}\theta^{*}$ as a function of $E_{\rm i}$.  Three inputs are thus required to perform the desired numerical calculation of $dR/dE_{\rm R}$.

\begin{figure}[t]
\centering
\includegraphics[height=1.42in]{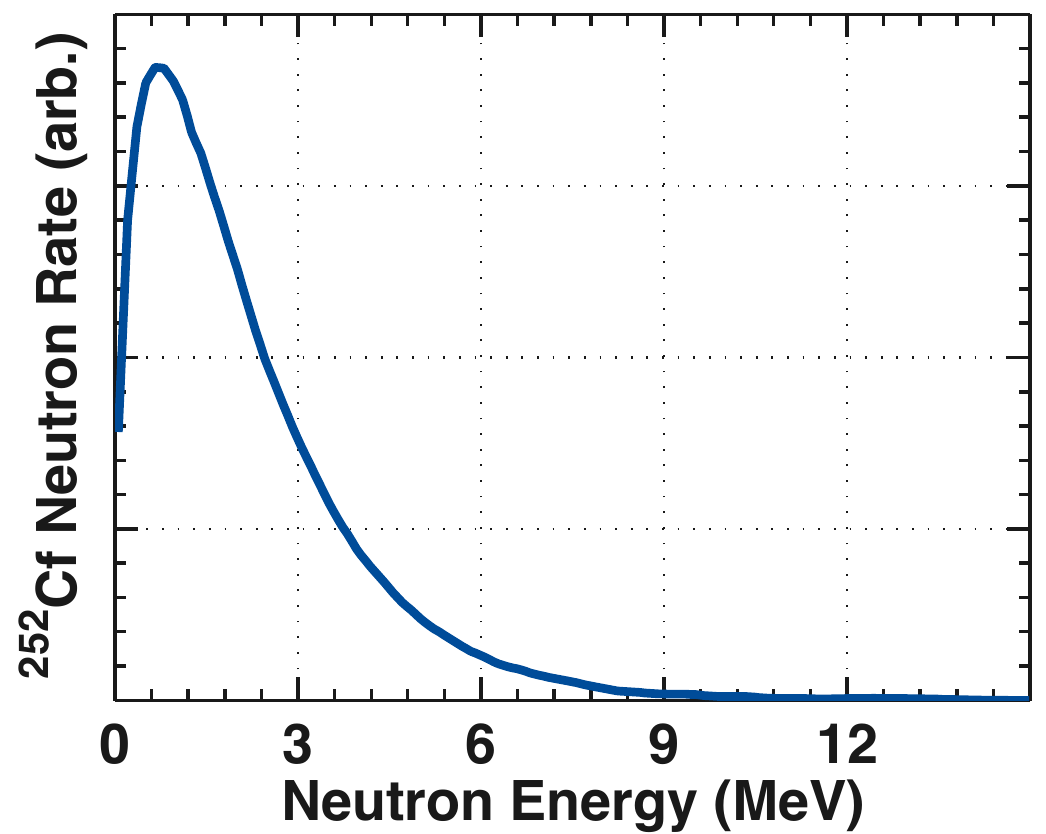} 
\includegraphics[height=1.42in]{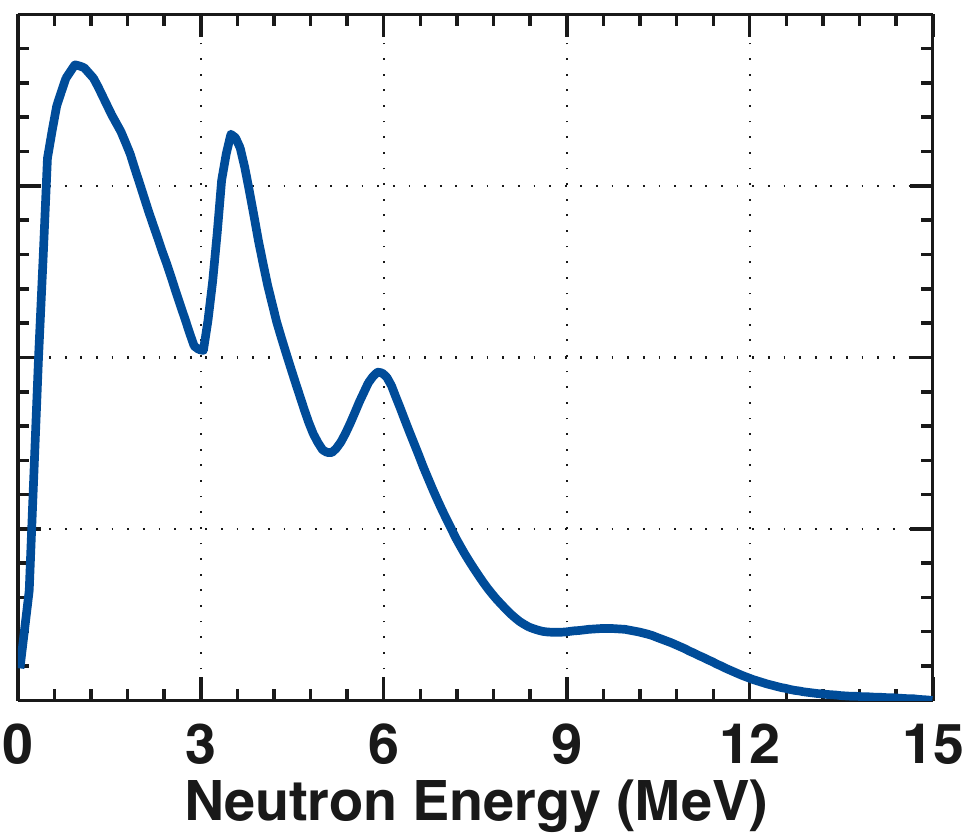} 
\caption{Spectra used in simulations of the CDMS shallow- (left) and deep-site (right) 
shielding configurations to represent the distribution of neutron energies emitted by $^{252}$Cf.  Figure adapted from~\cite{rbunkerthesis}.}
\label{fig:appA1}
\end{figure}

\subsection{Differential number density}
\label{app:A-2}

There appears to be some uncertainty regarding the high-energy tail of the spectrum of neutron energies emitted by $^{252}$Cf.  The distribution used for the CDMS shallow-site \oldgeant\cite{Geant3} simulations described in~\cite{kamat} (and used in~\cite{r21lowmass}) is approximately given by
\begin{equation}\label{eq:nrap7}
\frac{dn}{dE_{\rm i}} \propto e^{-E_{\rm i}/(1.42\textrm{\,MeV})}\sqrt{E_{\rm i}},
\end{equation}
and is shown in the left panel of Fig.~\ref{fig:appA1}.  
A complicated multi-peaked spectrum was used for the \geant 
simulations described in this paper and is shown in 
Fig.~\ref{fig:appA1}. Fortunately, the presence (or lack) of the high-energy structure exhibited by the deep-site spectrum does not appear to significantly affect the ZIP detector's nuclear-recoil response for recoil energies $<$100\,keV.  We explicitly confirmed with our \geant simulation that starting from either $^{252}$Cf spectrum yields differential event rates that are indistinguishable for recoil energies from a few to 100\,keV.

\begin{figure}
\centering
\includegraphics[width=\linewidth]{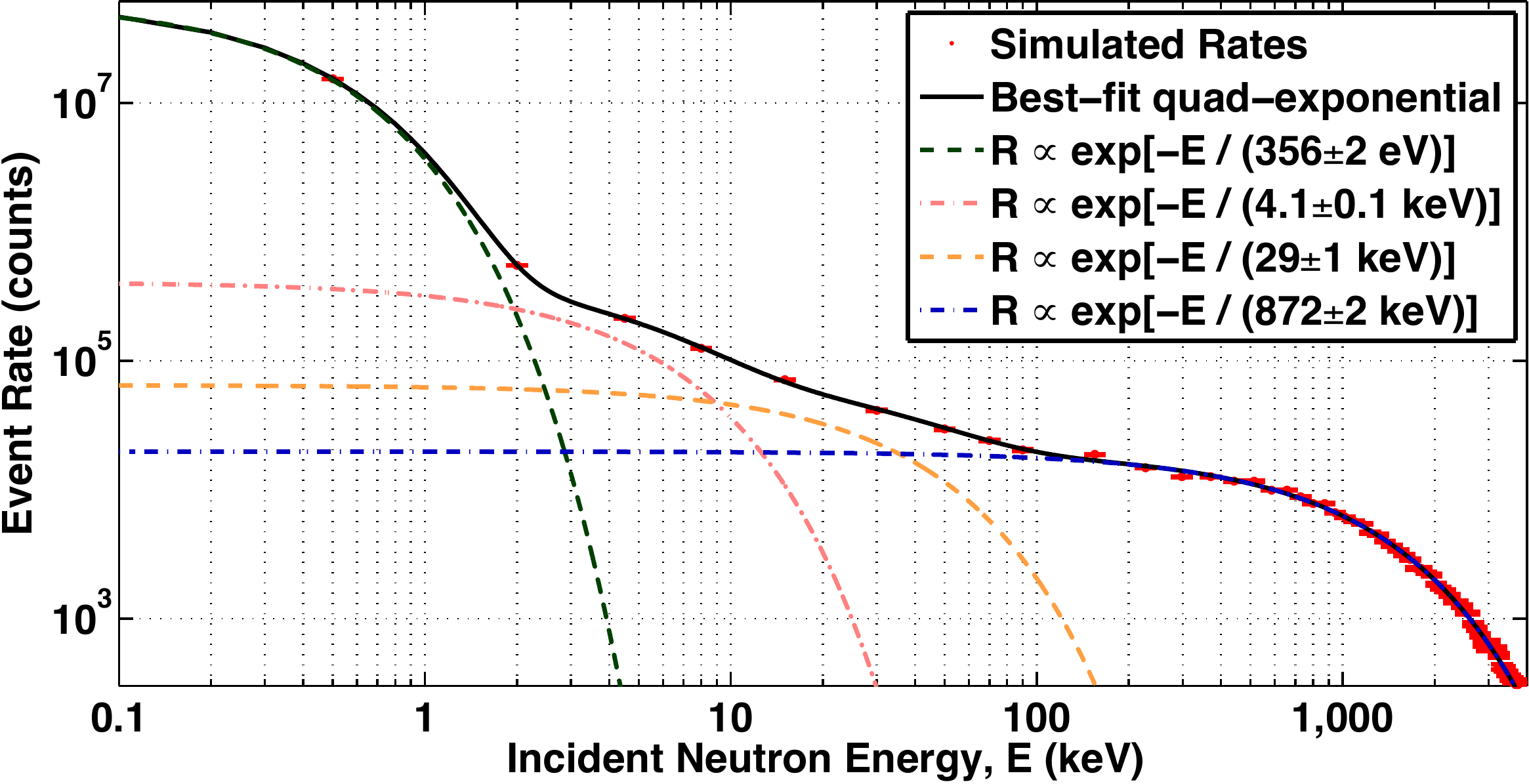} 
\caption{Spectrum of $^{252}$Cf neutron energies (dots with error bars) incident upon the ZIP detectors following simulated transport through the CDMS~II shield.  The multi-exponential fit (solid curve), consisting of components with characteristic energies 356\,eV (dark green dash), 4.1\,keV (light red dot-dash), 29\,keV (orange dash), and 872\,keV (blue dot-dash),  is used to evaluate $dR/dE_{\rm R}$ numerically.}
\label{fig:appA2}
\end{figure}

The spectrum of neutron energies directly emitted by the source is not quite what is needed for the numerical calculation. Neutron calibrations were typically conducted with the $^{252}$Cf source located such that the neutrons had to penetrate several layers of shielding in order to scatter from a ZIP detector. Consequently, the emitted energy spectrum must be transported through the CDMS~II shield to obtain the differential number density required by Eq.~\ref{eq:nrap5}.  We use the results of our \geant simulation rather than attempt a quasi-analytic estimate of this part of the calculation.  Simulated transport of the multi-peaked spectrum in Fig.~\ref{fig:appA1} through the CDMS~II shield yields the distribution shown in Fig.~\ref{fig:appA2}.  For convenience, this ZIP-incident spectrum is modeled by a multi-exponential fit.  The exponential with the largest decay constant contributes most of the events observed in the ZIP detectors and is given approximately by
\begin{equation}\label{eq:nrap8}
\frac{dn}{dE_{\rm i}} \propto e^{-E_{\rm i}/872\pm2\textrm{\,keV}}.
\end{equation}
The best-fit, eight-parameter (4 decay plus 4 normalization constants) multi-exponential indicated in Fig.~\ref{fig:appA2} provides the first input needed to evaluate $dR/dE_{\rm R}$ numerically.

\subsection{Elastic-scattering cross section}
\label{app:A-3}

The differential cross section for neutrons to elastically scatter from nuclei is composed of two parts.  The first part, denoted $\sigma(E_{\rm i})$, describes the dependence of the cross section on the incident neutron energy.  \geant~uses nuclear cross-section data from the \scshape endf \normalfont database to model $\sigma(E_{\rm i})$. For technical reasons, it was simpler to extract this information from the {\scshape jendl} database~\cite{jendl}. 
The {\scshape jendl} and {\scshape endf} databases for $^{28}$Si contain some very slight differences of $\sigma(E_{\rm i})$ for incident neutron energies greater than a few MeV. Additionally, the {\scshape jendl} cross sections cut off at 20\,MeV, while the {\scshape endf} cross sections extend to $\sim$150\,MeV.  
None of these differences are expected to significantly affect the recoil-energy spectra for $E_{\rm R} < 100$\,keV;  most events in the recoil-energy range of interest correspond to incident neutron energies less than a few MeV.

Due to the 20\,MeV limitation of the {\scshape jendl} cross sections, the evaluation of Eq.~\ref{eq:nrap5} was restricted to energies $<$20\,MeV.  Consequently, relative to the Monte Carlo simulated recoil-energy spectra presented earlier in this paper, the numerical calculation excludes a range of incident neutron energies between 20 and 150\,MeV (as well as any inelastic interactions).  The contribution to the differential event rate due to high-energy neutrons falls off exponentially with increasing energy, as does the number density per keV of incident energy.  The spectral shapes presented below are therefore expected to be trivially affected by this exclusion of the highest-energy incident neutrons.

\begin{figure}[t!]
\centering
\includegraphics[width=0.6\linewidth]{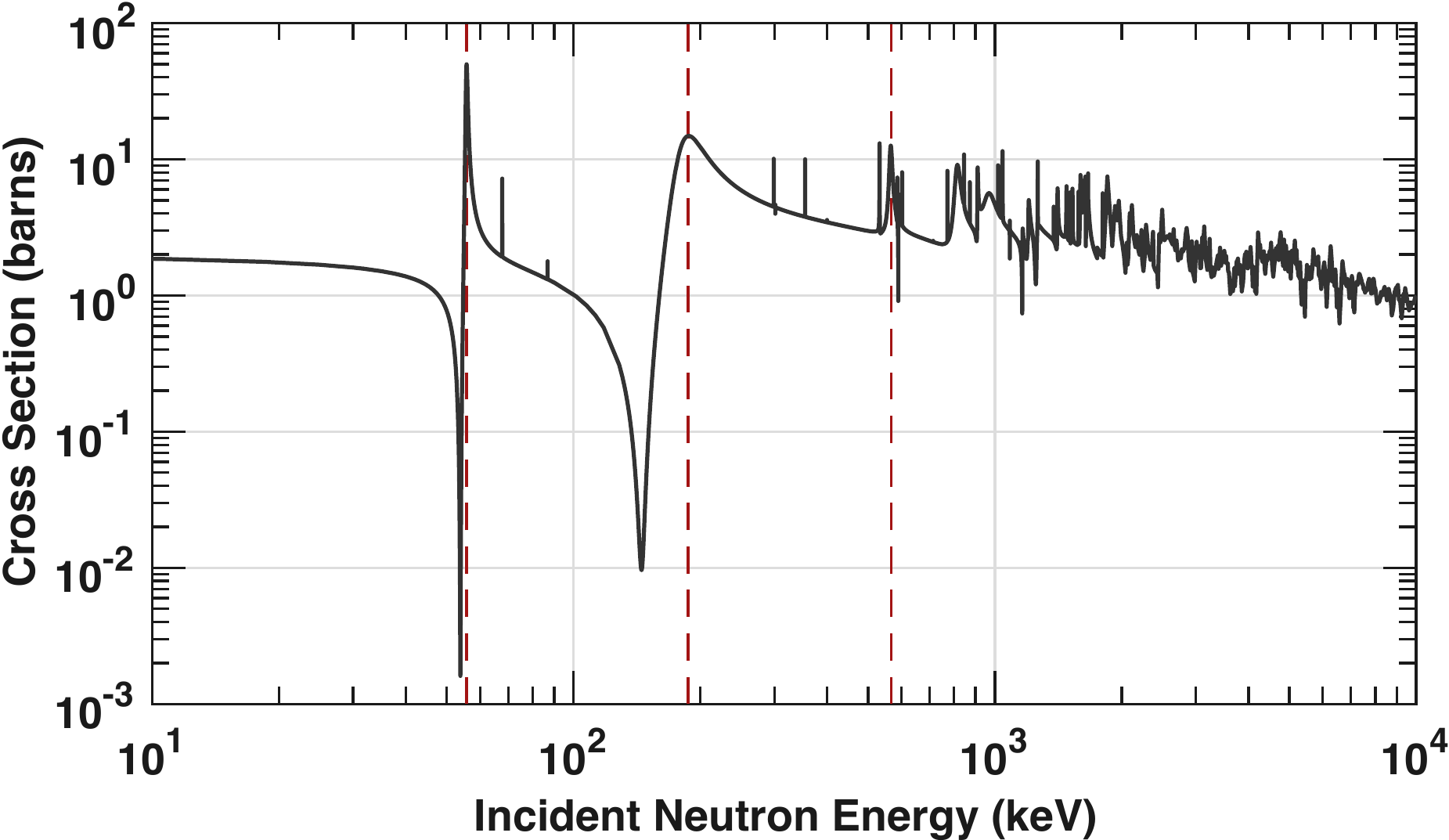}
\caption{Cross section for neutrons to elastically scatter 
from a $^{28}$Si target as a function of incident neutron energy. 
The cross section is interpolated from data 
in the \scshape jendl \normalfont database~\cite{jendl}.
The cross section exhibits 3 prominent resonances (red dashed) 
at $E_i \approx 55$, 183, and 550\,keV.}
\label{fig:appA3}
\end{figure}

The {\scshape jendl} database files are available as text files in which $\sigma(E_{\rm i})$ is listed at several discrete energies between 1$\times$10$^{-5}$\,eV and 20\,MeV.  To estimate $dR/dE_{\rm R}$ to the desired precision, it was necessary to interpolate between these discrete values such that $\sigma(E_{\rm i})$ could be evaluated at arbitrary energies.  The resulting interpolated cross sections for $^{28}$Si are shown in Fig.~\ref{fig:appA3}.

\subsection{Elastic-scattering angular probabilities}
\label{app:A-4}

The second part of the differential cross section, denoted $P(\textrm{cos\,}\theta^{*}|E_{\rm i})$, is the probability for a neutron of a given incident energy to scatter with a particular center-of-mass scattering angle.  These angular probabilities are stored in the \scshape endf \normalfont database files as Legendre polynomial coefficients.  Coefficients are provided at several discrete energies between 1$\times$10$^{-5}$\,eV and 150\,MeV, and can be used to construct the angular probabilities according to
\begin{equation}\label{eq:nrap9}
P(\textrm{cos\,}\theta^{*}|E_{\rm i}) = \frac{1}{2} + \sum_{l=1}^{N}\frac{2l+1}{2}a_{l}(E_{\rm i})\mathcal{P}_{l}(\textrm{cos\,}\theta^{*}),
\end{equation}
where $\mathcal{P}_{l}$ is the $l^{\textrm{th}}$ Legendre polynomial, $a_{l}(E_{\rm i})$ is the $l^{\textrm{th}}$ coefficient for incident energy $E_{\rm i}$, and the sum runs from $l=1$ to the highest-order nonzero term.  If there are no nonzero coefficients at a given incident energy, the cross section is isotropic (\textit{i.e.}, all angles are equally likely).  Similar to $\sigma(E_{\rm i})$, interpolation was used to obtain the angular probabilities at arbitrary energies.

\begin{figure}[t!]
\centering
\includegraphics[width=0.7\linewidth]{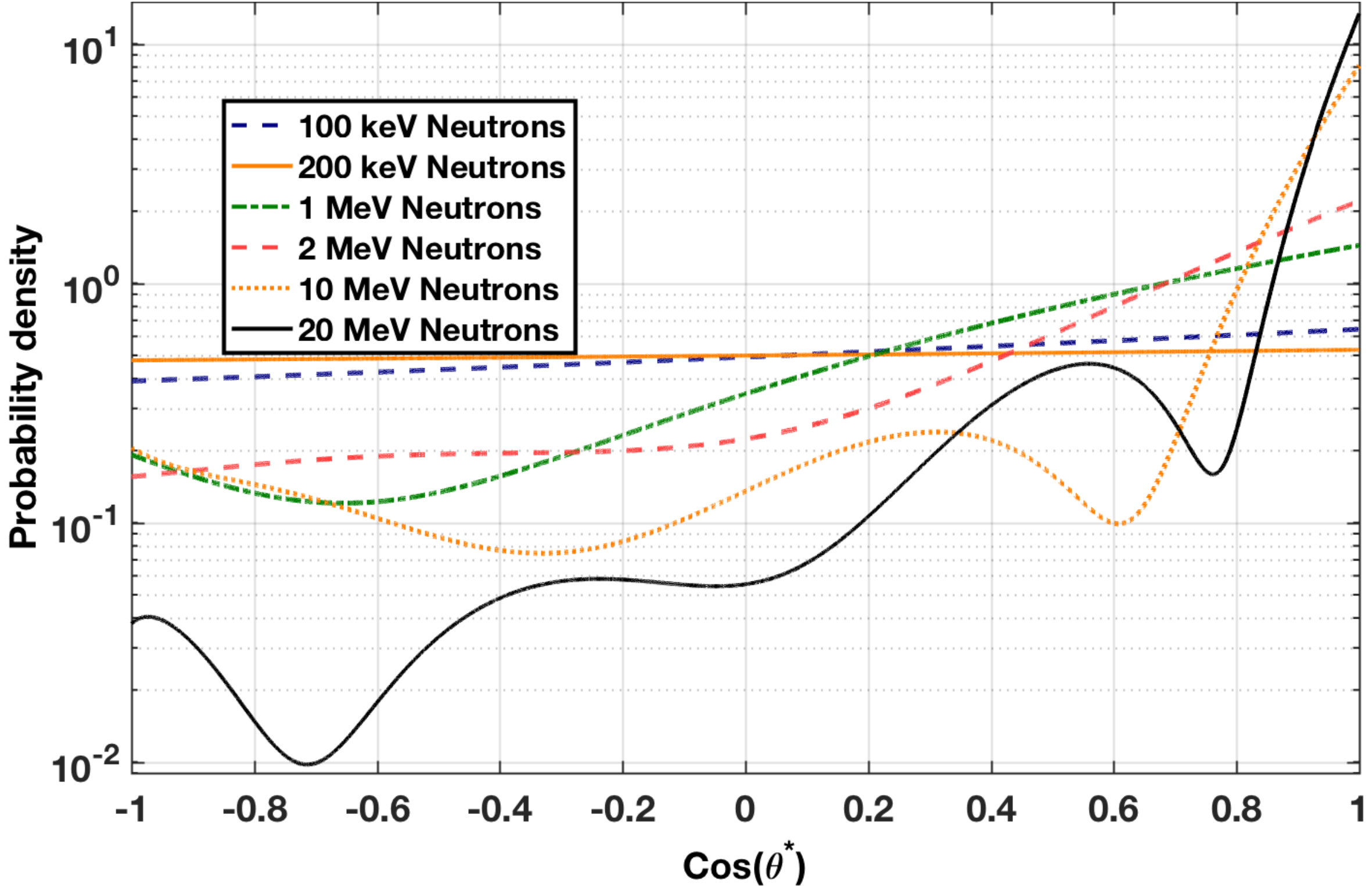}
\caption{Angular probability density for neutrons to scatter from $^{28}$Si for several values of incident neutron energy: 100\,keV (blue dash), 200\,keV (orange solid), 1\,MeV (green dot-dash), 2\,MeV (red dash), 10\,MeV (orange dash), 20\,MeV (black solid).  Angular data taken from the \scshape endf \normalfont database~\cite{endf}.}
\label{fig:appA5}
\end{figure}

The angular scattering probability densities are provided for multiple incident neutron energies 
in Fig.~\ref{fig:appA5}.
As the incident neutron energy increases, forward scattering ($\textrm{cos\,}\theta^{*} = 1$) becomes increasingly likely.

\subsection{Spectral shapes}
\label{app:A-5}

With the differential number density and cross-section data specified as described above, Eq.~\ref{eq:nrap5} was evaluated for $^{28}$Si for $E_{\rm R} = 1$--100\,keV in steps of 0.1\,keV.  At each recoil energy considered, a range of incident neutron energies was calculated (between $(1+A)^{2}E_{\rm R}/4A$ and 20\,MeV) as a function of $\cos\theta^*$. 
The three inputs described above were either 
evaluated or interpolated at each incident neutron energy, 
to determine the differential event rate at each $E_{\rm R}$.

\begin{figure}
\centering
\includegraphics[width=0.6\linewidth]{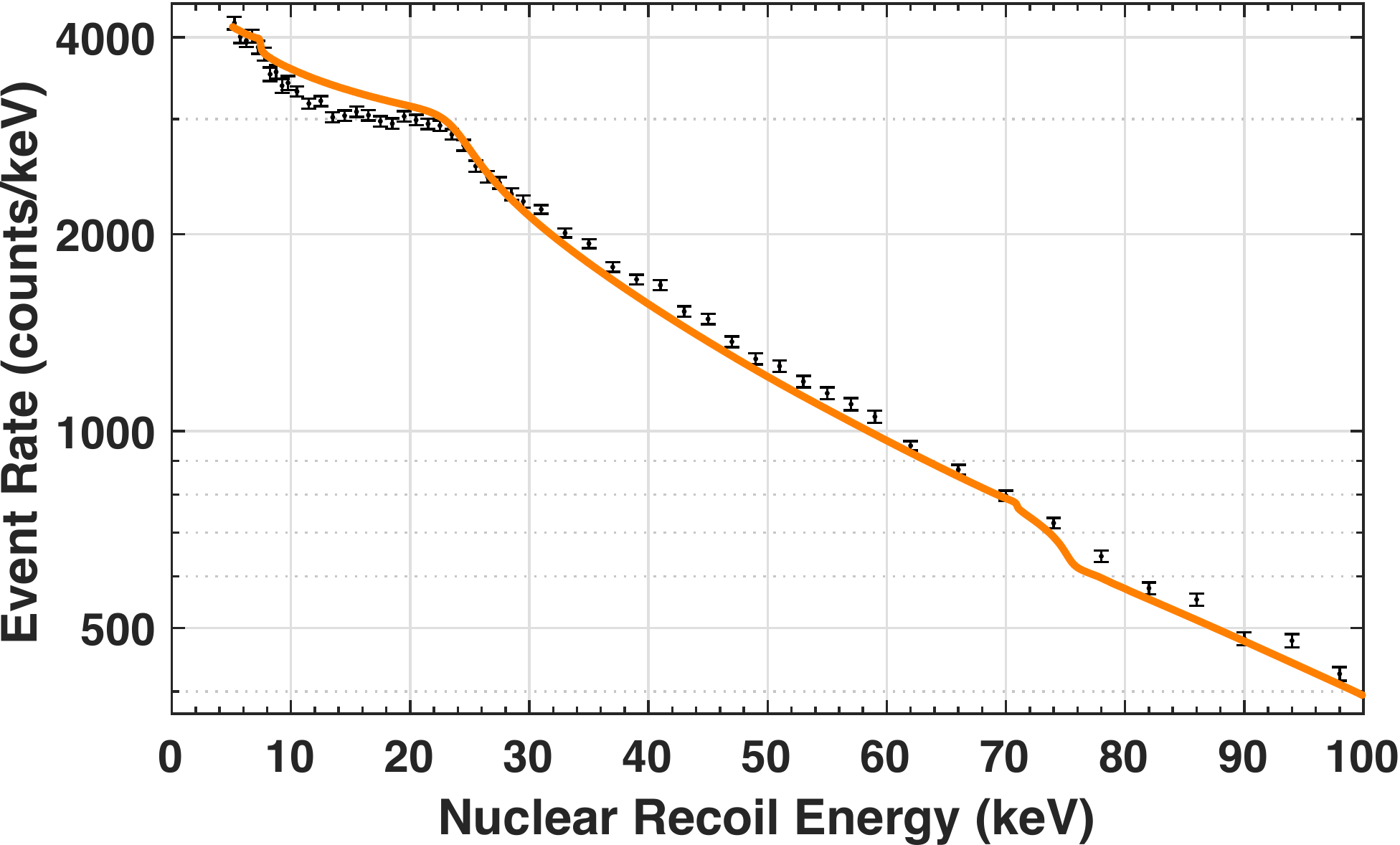} 
\caption{(Log-scale) Comparison of the \geant~\cite{geant403} simulated $^{252}$Cf nuclear-recoil event rate (black error bars) for the CDMS~II silicon detector ensemble with a numerical estimate of the spectral shape expected from $^{252}$Cf neutrons elastically scattering from a silicon target (line), where the former is given in total counts per keV and the latter is scaled to match the total integrated rate from 5 to 100\,keV.  The disagreement below $\sim$20\,keV is likely due to a combination of inelastic interactions and neutrons that multiply scatter in a single detector, effects included in the simulation but not in the numerical estimate.  In addition to the prominent feature near 20\,keV, there are smaller resonant features at $\sim$8 and 72\,keV.}
\label{fig:appA7}
\end{figure}

The resulting $^{28}$Si spectrum exhibits three bumps, shown in Fig.~\ref{fig:appA7}, 
due to the three most prominent resonances in the $^{28}$Si cross section 
(at $E_{\rm i} \approx 55$, 183, and 550\,keV; {\it cf.}\ Fig.~\ref{fig:appA3}). 
Figure~\ref{fig:appA7} also compares the numerically calculated silicon spectrum to the high-statistics \geant simulation results.  
The discrepancy at low energy might be related to inelastic interactions and the tendency for neutrons to multiply scatter (in a single detector), neither of which was taken into account by the numerical calculation.

\bibliographystyle{unsrt}
\bibliography{SI_NRS} 
\end{document}